\documentclass[a4paper,onecolumn,11pt,accepted=2026-01-16]{quantumarticle}
\pdfoutput=1
\usepackage{float}
\usepackage{graphicx}
\usepackage[svgnames]{xcolor}
\usepackage{hyperref}
\usepackage[numbers]{natbib}

\usepackage{amsthm,amsmath,amssymb,amsbsy}
\usepackage{bbm}
\usepackage{physics}
\usepackage{siunitx}
\usepackage{caption}
\usepackage{subcaption}

\newcommand{\mc}[1]{\mathcal{#1}}
\newcommand{\ox}{\otimes}
\newcommand{\id}{\mathbbm{1}}
\newcommand{\kb}{k_\text{B}}

\newcommand{\extn}{N_\text{ext}}
\newcommand{\entn}{N_\text{ent}}
\newcommand{\xzp}{\Delta X^{\text{zp}}_{\vb*{k}}}
\newcommand{\vari}{\mathrm{Var}}

\begin{document} 

\title{Macroscopic quantum coherence and entanglement in mechanical systems}

\author{Benjamin Yadin}
\email{benjamin.yadin@uni-siegen.de}
\affiliation{Naturwissenschaftlich-Technische Fakult\"at, Universit\"at Siegen, Walter-Flex-Stra\ss e 3, Siegen 57068, Germany}

\author{Matteo Fadel}
\email{matteofadel@pku.edu.cn}
\affiliation{Institute of Quantum Science and Technology $\&$ Frontiers Science Center for Nano-Optoelectronics, School of Physics, Peking University, Beijing 100871, China}
\affiliation{Department of Physics, ETH Z\"{u}rich, Z\"{u}rich 8093, Switzerland}

\begin{abstract}
    The notion of a macroscopic quantum state must be pinned down in order to assess how well experiments probe the large-scale limits of quantum mechanics.
    However, the issue of quantifying so-called quantum macroscopicity is fraught with multiple approaches having varying interpretations and levels of computability and measurability.
    Here, we introduce two measures that capture independent macroscopic properties: i) extensive size, measuring the degree of coherence in position or momentum observables relative to atomic-scale units, and ii) entangled size, quantifying the number of entangled subsystems.
    These measures unify many desirable features of past proposals while having rigorously justified interpretations from quantum information theory in terms of coherence and multipartite entanglement.
    We demonstrate how to estimate them and obtain lower-bounds using experimental data from micro-mechanical oscillators and diffracting molecules.
    Notably, we find evidence for genuine multipartite entanglement of $10^5$ to $10^7$ atoms in recent mechanical superposition states.
\end{abstract}

\maketitle

\section{Introduction}

Can quantum behaviour extend to the macroscopic scale?
Schr\"odinger's cat was born in a thought experiment highlighting the paradoxical consequences of macroscopic quantum coherence~\cite{Schrodinger1935Die}.
While there is nothing in the theory of quantum mechanics itself that prevents this in principle, it is never seen in practice.
One good explanation is that unwanted interactions with an environment inevitably cause decoherence, whereby quantum properties of a system are diluted over time -- and this typically occurs faster for larger systems \cite{Zurek2003Decoherence}.
It is also often argued that the precision resources for detecting quantum behaviour are highly demanding for macroscopic systems \cite{Peres1964Macroscopic,Kofler2007Classical,Raeisi2011Coarse,Sekatski2014Size,Skotiniotis2017Macroscopic}.
Taking such a viewpoint, one would expect macroscopic quantum states to be attainable once the practical limitations are overcome.
However, new and currently untested physical phenomena, such as effects from quantum gravity, could be responsible for unavoidable collapse of the wavefunction.
It is the role of experiments to help settle this question by probing ever-larger quantum regimes.
A notable class of systems making headway in this direction in recent years is those involving mechanical degrees of freedom, including micro-mechanical oscillators and diffracting molecules.

However, the question of what counts as a macroscopic quantum state -- or how to quantify the degree of macroscopic quantumness -- poses conceptual and technical challenges.
From the outset, one has to define what is meant by both macroscopic and quantum.
Neither of these terms has a unique meaning: ``macroscopic'' could reasonably refer to quantities at the human scale (around the metre, second, or kilogram scale) or to a collection of many particles around the Avogadro number; further, ``quantumness'' is these days understood to manifest in a large variety of phenomena, the best known being superposition and entanglement.
One hopes to have a computable and experimentally obtainable measure that indicates in some way the extent to which quantum behaviour is present at a large scale, with higher numbers indicating an approach towards a truly macroscopic quantum state.
An array of quantum \textbf{macroscopicity} measures have been devised in a range of approaches~\cite{Frowis2018Macroscopic}.
Some examples include using the size of quantum fluctuations~\cite{Shimizu2005Detection,Lee2011Quantification}, the scale of quantum superpositions~\cite{Cavalcanti2008Criteria,Yadin2016General}, distillation of multipartite entanglement~\cite{Dur2002Effective,Yadin2015Quantum}, single-particle operations or measurements~\cite{Korsbakken2007Measurement,Marquardt2008Measuring}, and sensitivity to decoherence from minimal modifications of the Schr\"odinger equation~\cite{Nimmrichter2013Macroscopicity,Schrinski2019Macroscopicity,Schrinski2023Macroscopic}.
Although all aim in some way to capture the degree of resemblance to an ideal ``cat state'', they are often hard to compare and have varying scopes of applicability to different types of systems or specific families of states.

Here, our goal is to bring some clarity at least to the domain of mechanical systems.
We introduce two measures that capture two primary ways of being macroscopically quantum.
One can ask of a quantum state: i) over what physical length or momentum scales does quantum coherence extend, and ii) are the observable properties only explainable by quantum mechanics operating simultaneously on many particles?
Our measures correspond to Leggett's key concepts of extensive difference and disconnectivity~\cite{Leggett1980Macroscopic}, and are justified by a solid foundation on modern quantum information theory to make their interpretation transparent and to carefully characterise their properties.
The construction follows the approach to macroscopicity based on quantum Fisher information~\cite{Frowis2012Measures,Oudot2015Two,Yadin2016General} but employs two different normalisations to form the measures.
Further, they have the advantages of being well-defined for all quantum states, easily computable, and possible to estimate directly from experimental data.

In Sec.~\ref{sec:measures}, we explain the motivation for our approach, define the measures, and establish their properties.
Next, in Sec.~\ref{sec:leggett_example}, we illustrate the measures in a thought experiment by Leggett involving a quantum superposition at the boundary of the macroscopic scale.
We then perform estimates for several experiments with mechanical oscillators and diffracting molecules in Secs.~\ref{sec:oscillator} and \ref{sec:diffraction} respectively.
Finally, in Sec.~\ref{sec:conclusions}, we summarise the findings and provide an outlook on extensions to a wider range of systems.

\section{Measures}
\label{sec:measures}

\subsection{Rationale} \label{sec:rationale}
As argued by Leggett~\cite{Leggett2002Testing}, there is unlikely to be a single measure that fully captures the macroscopicity of a quantum state.
However, there are two main properties that clearly must be ``large'' for a quantum state to be macroscopic.
These can be illustrated by considering an ideal Schr\"odinger cat state.
Firstly, the cat is in a superposition $\ket{\psi_0}+\ket{\psi_1}$ of two branches that differ appreciably in some macroscopic observable.
For example, one could distinguish the branches by a measurement of the cat's centre of mass $X_\text{CM}$: in $\ket{\psi_0}$, the cat is lying down, and in $\ket{\psi_1}$ it is standing up, so the value of $X_\text{CM}$ differs on a macroscopic (i.e., ``human'') scale, say $\Delta X_\text{CM} = \SI{.1}{\metre}$.
In this example, Leggett's \textbf{extensive difference} is found by forming the extensive\footnote{An extensive quantity is one that is additive with respect to subsystems.} quantity $Q = M X_\text{CM}$, where $M$ is the total mass, and expressing it in some atomic-scale units $Q_0$.
We take $Q_0 = m_u a_0$ (the unified atomic mass unit times the Bohr radius), giving $\Delta Q/Q_0 = M \Delta X_\text{CM}/Q_0 \sim 10^{36}$ for the superposition.

Secondly, the cat is composed of a number of atoms of order $10^{26}$, so in the superposition state, roughly this many particles are entangled.
A tentative measure of the number of entangled particles was proposed by Leggett in Ref.~\cite{Leggett1980Macroscopic}, termed the \textbf{disconnectivity}, however he states ``from the point of view of modern quantum information theory\dots [it] could almost certainly be substantially improved''~\cite{Leggett2002Testing}.
Such a quantity is needed to distinguish a Schr\"odinger-cat situation from something like a single neutron passing through an interferometer whose branches are $\SI{0.1}{\metre}$ apart, having a still large extensive difference $\Delta Q/Q_0 \sim 10^9$.

Here, we propose two measures that capture these properties, termed the \textbf{extensive size} and \textbf{entangled size}.
They have the advantages of being well-defined for all quantum states -- beyond only pure-state superpositions of two branches -- and of being possible to estimate directly from experimental data.

\subsection{Quantum Fisher information}

The two proposed measures are based on \textbf{quantum Fisher information} (QFI).
This quantity is central to quantum metrology, where it determines the minimal uncertainty with which one can estimate a parameter encoded into a probe state.
For our purposes, we consider a given state $\rho$ and observable $A$, with the parameter $\theta$ encoded via unitary evolution generated by $A$ as
\begin{align}
    \rho_\theta := e^{-i\theta A} \rho e^{i\theta A}.
\end{align}
The quantum Cram\'er-Rao bound states that, with measurements on $n$ copies of $\rho_\theta$, the minimum possible uncertainty in an estimate of $\theta$ is lower-bounded by~\cite{Braunstein1994Statistical}
\begin{align}
    \delta \theta \geq \frac{1}{\sqrt{n\, \mc{F}(\rho, A)}}.
\end{align}
This bound can be achieved in the limit of many copies.
Here, $\mc{F}(\rho,A)$ is the QFI of $\rho$ with respect to $A$; one definition is given by the fidelity $F(\rho,\rho_\theta) = \tr\sqrt{\sqrt{\rho}\rho_\theta \sqrt{\rho}}$ between $\rho$ and $\rho_\theta$ for small $\theta$:
\begin{align}
    F(\rho,\rho_\theta) = 1 - \frac{1}{8} \mc{F}(\rho,A) \theta^2 + \mc{O}(\theta^3).
\end{align}
The QFI therefore quantifies the (squared) speed with which $\rho$ changes under the unitary evolution.
Alternatively, it is the maximal \emph{classical} Fisher information that can be extracted from the probe state under any generalised measurement (i.e., any POVM $\{E_i\}$ with $E_i \geq 0$ and $\sum_i E_i = \id$).
A given measurement on $\rho_\theta$ results in a parameter-dependent probability distribution $p_\theta^i = \tr[\rho_\theta E_i]$, for which the classical Fisher information is
\begin{align}
    \mc{F}_\text{cl}(\{p_\theta^i\}) = \sum_i \dfrac{1}{p_\theta^i} \left( \dfrac{\partial p_\theta^i}{\partial\theta} \right)^2.
\end{align}
In general, one has $\mc{F}_\text{cl}(\{p_\theta^i\}) \leq \mc{F}(\rho,A)$, and there is always a choice of measurement that makes them equal.
It is worth noting that the QFI reduces to (four times) the variance of $A$ when the state is pure: $\frac{1}{4}\mc{F}(\dyad{\psi},A) = \vari (\dyad{\psi},A) := \ev{A^2}{\psi} - \ev{A}{\psi}^2$.
\\

It is not self-evident that the utility of a probe quantum state for metrology should be connected with macroscopicity.
This insight was developed over a series of works including Refs.~\cite{Shimizu2002Stability,Shimizu2005Detection,Frowis2012Measures,Frowis2015Linking,Oudot2015Two} by considering the information-theoretical properties of the QFI and its behaviour on certain families of states and observables.
In general, $A$ is taken to be some \textbf{extensive observable}: for $N$ constituent parts, typically of the form $A = \sum_{i=1}^N A_i$, where $A_i$ acts only on part $i$.
This is sometimes also called a collective observable, and one is typically allowed to maximize the QFI over all observables within this class, in order to pick out the most relevant degree of freedom.
In Ref.~\cite{Frowis2012Measures} it was noted that the QFI of an $N$-qubit state with $A = \sum_i \sigma^\alpha_i$ (where $\sigma^\alpha$ is a Pauli matrix) acts as a witness of genuine many-particle entanglement.
Moreover, the QFI is largest, with the value $4N^2$, for a ``cat state'' of the form $\ket{0}^{\ox N} + \ket{1}^{\ox N}$ (up to a change of local basis).
The ``Heisenberg scaling'' limit of quantum metrology, where $\mc{F} \propto N^2$, is taken as indicative of a family of macroscopic quantum states, compared with $\mc{F} \propto N$ for unentangled states.
In other systems, different classes of observables are chosen to pick out the desired macroscopic quantum states: for example, other collective operators in higher-dimensional discrete systems, or position and momentum in phase space for continuous-variable systems~\cite{Frowis2015Linking,Oudot2015Two}.

However, it has been unclear so far how to correctly normalise the QFI to obtain a quantity that has a meaningful interpretation and can be used to compare different systems.
Below, we motivate two normalisations to define measures that pick out important macroscopic quantum properties.

\subsection{Extensive size}

The first measure quantifies the amount of coherence in the state with respect to the eigenbasis of the macroscopic observable.
A good rigorously motivated measure of coherence is the QFI, as shown in Ref.~\cite{Yadin2016General} from a resource-theoretical perspective.
In brief, this is justified by it satisfying some important properties.
Firstly, $\mc{F}(\rho,A) = 0$ if and only if $\rho$ is incoherent, i.e., has vanishing off-diagonal elements in the eigenbasis of $A$.
More precisely, if $A = \sum_{a,g} a \dyad{a,g}$, where $a$ label distinct eigenvalues and $g$ labels degeneracies, then $\mc{F}(\rho,A) = 0$ when $\mel{a,g}{\rho}{a',g'}=0$ for any $a \neq a'$.
Secondly, $\mc{F}(\rho,A)$ cannot be increased by any process that does not inject additional coherence into the system.
Such processes are defined formally as \textbf{covariant channels} -- those that commute with the unitary evolution generated by $A$.
Any covariant channel $\mc{E}$ can be realised as a physical process where the system interacts with an environment that is incoherent with respect to some observable $A'$, under a global unitary evolution that conserves the total quantity $A + A'$.
In terms of coherence, covariant channels have the crucial property that a matrix element $\ketbra{a,g}{a',g'}$ evolves onto other matrix elements that have the same difference $a-a'$.
This ensures that the physical scale of a superposition between branches with values $a$ and $a'$ is retained, but it may decohere.
If $A$ is a macroscopic observable, then we have the following interpretation:
since macroscopicity is associated with coherence corresponding to macroscopic values of $a-a'$, this property ensures that covariant channels can only degrade the macroscopicity of a state relative to $A$.\\

The QFI actually has a unique meaning as the \textbf{coherence cost} of preparing a state.
This was proven in Ref.~\cite{Marvian2022Operational} for the case of qubits.
Consider starting from a number $n$ of coherence ``units'' $\ket{\phi} = (\ket{0}+\ket{1})/\sqrt{2}$, with the aim of preparing as many copies as possible of the target state $\rho$ via a covariant channel.
Here, we take $A = \sum_i \sigma^z_i$, and the observable is added over multiple copies.
One can obtain $m$ copies of $\rho$ with vanishing error in the limit of large $n$, at a rate $m / n = \mc{F}(\dyad{\phi},A) / \mc{F}(\rho,A)$.
Thus, the coherence cost of $\rho$ is $\mc{F}(\rho,A)/\mc{F}(\dyad{\phi
},A)$, given $\phi$ as a unit.
While a rigorous mathematical statement has only been proven for qubits, we extend the same interpretation to generic systems (for which a weaker version of the statement is at least known with pure states~\cite{Yadin2016General,Yadin2017Thesis}).\\

We then define the \textbf{extensive size} of $\rho$ relative to $A$ as
\begin{align}
    \extn(\rho,A) := \frac{\mc{F}(\rho,A)}{4 A_0^2},
\end{align}
where $A_0$ is some chosen atomic-scale unit.
Here, we are concerned with mechanical systems and therefore focus on $A$ being one of position or momentum as preferred observables.
For a collection of $N$ particles with masses $m_i$, positions $x_i$ and momenta $p_i$, we typically use the position of the centre-of-mass multiplied by the total mass, and momentum:
\begin{align}
    Q & := \sum_{i=1}^N q_i = \sum_{i=1}^N m_i x_i = M X_\text{CM}, \\
    P & := \sum_{i=1}^N p_i,
\end{align}
where $M = \sum_i m_i$.
(In Sec.~\ref{sec:oscillator_basics}, we extend this approach to higher overtone modes of an oscillator.)
To set atomic-scale units, we take $\ket{\phi}$ to represent a particle of mass $m_u$ (atomic mass unit) with a minimal-uncertainty Gaussian wavefunction localised over a distance $a_0$ (the Bohr radius).
The corresponding normalisation units are then
\begin{align}
    Q_0 & = m_u a_0, \\
    P_0 & = \frac{\hbar m_u}{2Q_0} = \frac{\hbar}{2a_0}.
\end{align}

To see the connection with Leggett's extensive difference, consider the model of Schr\"odinger's cat discussed in Sec.~\ref{sec:rationale}, with an equal superposition of two branches each having negligible position uncertainty and a centre-of-mass difference $\Delta X_\text{CM}$, corresponding to $\Delta Q = M \Delta X_\text{CM}$.
Since $\rho$ is considered to be pure, we have $\extn(\rho,Q) = \vari (\rho,Q)/Q_0^2 = (\Delta Q/2Q_0)^2$ -- the \emph{square} of the  extensive difference.
This square relationship is crucial to ensure that $\extn$ is an extensive quantity in the sense of being additive for independent systems, since $\mc{F}(\rho_1 \ox \rho_2, A_1 + A_2) = \mc{F}(\rho_1,A_1) + \mc{F}(\rho_2,A_2)$.
In more familiar terms, the \emph{variance} of a distribution -- rather than the standard deviation -- is additive for independent systems.

We note that there is some relation with the macroscopicity measure $\mu$ of Ref.~\cite{Nimmrichter2013Macroscopicity}, which is defined in terms of the ability of an \emph{experiment} (rather than a single state) to rule out a range of parameters in a minimal modification to quantum mechanics.
In the case of a compact particle in a diffraction experiment, in a spatial superposition state $\rho$ whose coherence is maintained for a time $\tau$, we find that $\mu \sim \log_{10}[\extn(\rho,Q)] + \log_{10}[\tau / \SI{1}{\second}]$, see Appendix~\ref{app:nh}. 
This relation derives from the QFI implying a sensitivity to the models considered in Ref.~\cite{Nimmrichter2014Macroscopic}; it shows that the measure $\mu$ corresponds closely in a range of cases to the extensive size, but includes a contribution from the coherence time achieved in the experiment.

\subsection{Entangled size}
The second measure aims instead to capture multipartite correlations between parts of a system.
This is therefore defined not only relative to a chosen observable $A$, but also to a chosen \textbf{partition} $\Pi = \{\Pi_1,\Pi_2,\dots,\Pi_n\}$ of the system into $n$ subsystems.
The partition $\Pi$ must be such that $A = \sum_{i=1}^n A_{\Pi_i}$ is an additive observable with no overlapping terms between different subsystems $\Pi_i$.
The \textbf{entangled size} is then defined as
\begin{align} \label{eqn:entn_def}
    \entn(\rho,A,\Pi) := \frac{\mc{F}(\rho,A)}{4 \sum_{i=1}^n \vari (\rho,A_{\Pi_i})},
\end{align}
where the denominator is the sum over all variances of the local observables $A_{\Pi_i}$ acting on each subsystem in the partition.
Note that the choice of the partition $\Pi$ is arbitrary, and different choices can have a great impact on the value of $\entn$, as demonstrated below in Sec.~\ref{sec:leggett_example}.

The ability of $\entn$ to act as an entanglement witness was proven in Refs.~\cite{Yadin2017Thesis,Fadel2023Multiparameter}; here, we give a different proof via the following characteristic properties in Appendix~\ref{app:entn_properties}:

\begin{enumerate}
    \item \textbf{Maximum size:} For any partition with $n$ elements, 
    \begin{align} 
        \entn(\rho,A,\Pi) \leq n.
    \end{align}
    Moreover, the upper limit of $n$ is achieved if and only if $\rho$ has the form of a generalised GHZ state -- see Appendix~\ref{app:max_size} for details.
    For example, in a simple case where the subsystems are qubits, this form reduces to $\sqrt{1-q} \ket{0}^{\ox n} + \sqrt{q} \ket{1}^{\ox n}$, where $\ket{0}, \ket{1}$ are the eigenvectors of the local observables $A_{\Pi_i}$.

    \item \textbf{Independent systems:} For a product state $\rho_\alpha \ox \rho_\beta$, where $\alpha$ and $\beta$ represent two disjoint parts of the whole system,
    \begin{align}
        \entn(\rho_\alpha \ox \rho_\beta, A_\alpha + A_\beta, \alpha \cup \beta) \leq \max \{ \entn(\rho_\alpha,A_\alpha, \alpha),\, \entn(\rho_\beta,A_\beta, \beta) \}.
    \end{align}

    \item \textbf{Classical mixtures:} For any pair of states $\rho,\, \sigma$ of the same system, and mixing probability $p \in [0,1]$,
    \begin{align}
        \entn(p \rho + [1-p] \sigma, A, \Pi) \leq \max \{ \entn(\rho,A,\Pi) ,\, \entn(\sigma,A,\Pi) \}.
    \end{align}

    \item \textbf{Multipartite entanglement witness:} Suppose $\rho$ has no $(k+1)$-partite entanglement between subsystems $\Pi_i$.
    Formally, $\rho$ is $k$-producible~\cite{Guhne2005Multipartite}, meaning that it can be written as a classical mixture of products of the form $\ket{\psi_1} \ket{\psi_2} \dots$ in which each block $\ket{\psi_j}$ entangles no more than $k$ subsystems.
    (The blocks need not be consistent between different terms in the mixture.)
    Then we have
    \begin{align}
        \entn(\rho,A,\Pi) \leq k.
    \end{align}
\end{enumerate}

Informally, $\entn$ can therefore be interpreted as a quantifier of multipartite entanglement -- although it is strictly only a witness, in that a value greater than $k$ implies that the state contains genuine $(k+1)$-partite entanglement, but not the converse.
Property (1) shows how $\entn$ picks out the degree of resemblance to an entangled Schr\"odinger cat state.
It captures a particular kind of correlation, characteristic of a cat-type state, which is largest when subsystems are positively correlated (if all $A_{\Pi_i}$ are alike).
Properties (2) and (3) are important in showing that $\entn$ is never increased under either classical mixing, or (unlike $\extn$) under combining independent systems.\\

\section{Illustrative example} \label{sec:leggett_example}

In order to present the intuition for our measures, we consider a state introduced in a thought experiment by Leggett in Ref.~\cite{Leggett2016Note}, supposed to be just at the boundary of what one would naturally call macroscopic.
He considers a piece of solid matter -- for concreteness, a cube of the crystal LiF of size $\SI{5}{\micro\metre}$, about the smallest size resolvable by the human eye.
This is in a superposition $\ket{\psi_0} + \ket{\psi_1}$ where each branch is a ``classical'' state with negligible position and momentum uncertainty, moving relative to each other at a speed of $v = \SI{5}{\micro\metre\per\second}$.
Initially, with full spatial overlap, this should be regarded as a momentum superposition; after time $t=\SI{1}{\second}$, the branches are non-overlapping in position and so it also becomes a spatial superposition.
We focus first on the $t=0$ state.

For simplicity, we model the crystal as being composed of a number $N = \num{1.6e13}$ of a single species of atom of mass $m = 12.5 m_u$.
We ignore electrons in the calculations, as the greater mass of the nuclei means that they provide the dominant contribution.
With the momentum difference between branches as $\Delta P = Mv \approx \SI{1.7e-18}{kg.m.s^{-1}}$, the extensive size for momentum is $\extn(\rho,P) = (\Delta P / 2P_0)^2 \approx \num{6.9e11}$.

\begin{figure}[h]
    \centering
    \includegraphics[width=0.98\textwidth]{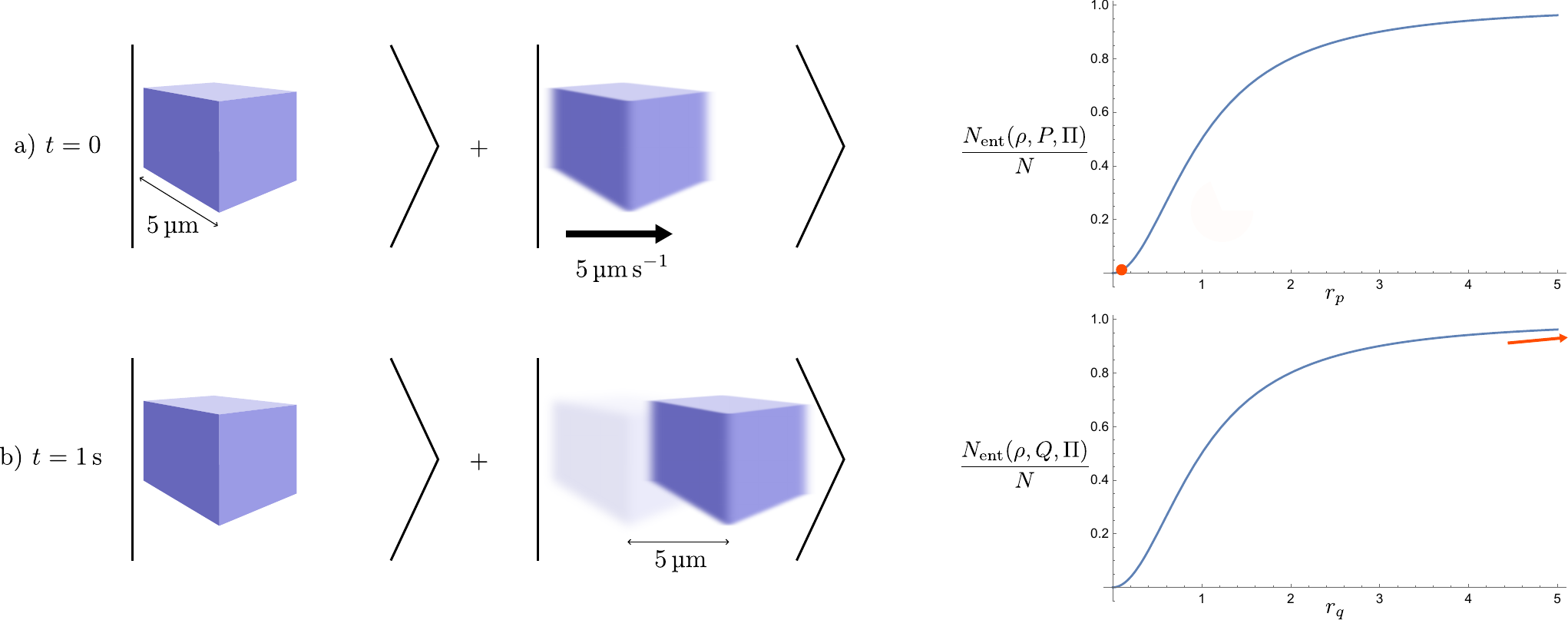}
    \caption{Illustration of Leggett's crystal example: a superposition of a $\SI{5}{\micro\meter}$-cubed LiF crystal in two total momentum states. a) Initially, the position of the centre-of-mass is the same in the two branches.
    The entangled size (as in Eq.~\eqref{eqn:crystal_entn}) of the momentum superposition is a small fraction of the atom number, due to the small ratio $r_p = \Delta p /2 \delta p$ of each atom's momentum shift $\Delta p$ between branches to its momentum uncertainty $\delta p$ in each branch.
    b) After $t=\SI{1}{\second}$, the branches are separated by $\SI{5}{\micro\meter}$ and the corresponding ratio $r_q$ is very large, implying high distinguishability of the branches and a position entangled size approaching the atom number.
    }
    \label{fig:crystal}
\end{figure}

For the entangled size, first consider the partition $\Pi$ into single atoms.
In the comoving frame of each classical branch, we take the crystal to be thermalised at room temperature with negligible uncertainty in its centre-of-mass momentum.
Then every atom is confined in the crystal to a region of approximate size $\delta x \sim \SI{e-11}{m}$ (see Ref.~\cite{Cartz1955Thermal} for typical values in solids at room temperature -- we discuss this in more detail in Sec.~\ref{sec:oscillator_basics}), with a corresponding single-atom momentum standard deviation $\delta p \sim \hbar/2\delta x \sim \SI{5.3e-24}{kg.m.s^{-1}}$.
In the superposition state, we include the uncertainty from the difference in single-atom momentum between the branches to get $\vari (\rho,P_i) \approx \delta p^2 + (\Delta p/2)^2$, where $\Delta p := \Delta P / N \approx \SI{1.0e-31}{kg.m.s^{-1}}$.
Then we obtain the useful expression
\begin{align} \label{eqn:crystal_entn}
    \entn(\rho,P,\Pi) & \approx \frac{(\Delta P/2)^2}{N [\delta p^2 + (\Delta p/2)^2]} = N \left(\frac{r_p^2}{1+r_p^2} \right), \nonumber \\
    \text{where } r_p & = \frac{\Delta p}{2\delta p}.
\end{align}
This identifies $r_p$ as an important parameter -- it is the ratio of the single-atom momentum difference between branches to the single-atom momentum uncertainty per branch\footnote{Note that the form of $\entn$ in Eq.~\eqref{eqn:crystal_entn} is similar to other macroscopicity calculations in past literature for models of cat states with branches that are not fully macroscopically distinct~\cite{Dur2002Effective,Bjork2004Size,Korsbakken2007Measurement}.}.
This quantifies the local distinguishability of the branches, as determined by a momentum measurement.
For small $r_p$, we have $\entn \approx N r_p^2$, with $\entn$ approaching the maximum value $N$ with increasing $r_p$.
In the present example, we have $r_p \approx \num{e-8}$, giving $\entn(\rho,P,\Pi) \approx N r_p^2 \sim \num{e-3}$.
Thus, it is not possible to witness \emph{any} entanglement in this state.
However, using the state $\rho_{\SI{1}{s}}$ after $t = \SI{1}{s}$ of free evolution (or even less, say $\SI{50}{\micro \second}$), the analogous calculation for position gives $r_q = \frac{\Delta q}{2\delta q} = \frac{\Delta X}{2\delta x} \sim \num{e5}$ so that $\entn(\rho_{\SI{1}{s}},Q,\Pi) \approx N \sim \num{e13}$ -- confirming our expectation that this state resembles a true macroscopic spatial superposition.
Moreover, this state has the huge value $\extn(\rho_{\SI{1}{s}},Q) \approx \num{8.9e37}$.

To see the importance of the partition choice, we compare the partition into atoms with that into individual nucleons.
The momentum uncertainty for single nucleons is much larger: as a rough estimate, since each nucleon is confined to a scale of $\delta x \sim \SI{e-15}{m}$, we have $\delta p \sim \hbar / (2\delta x) \sim \SI{5e-20}{kg.m.s^{-1}}$, thus $r_p$ is a factor of about $\num{e5}$ smaller (including the smaller momentum shift per nucleon), and $\entn(\rho,P,\Pi_\text{nucl})$ is about $\num{e9}$ times smaller.
Hence, atoms, rather than nucleons, are the right subsystems to choose in order to maximise the momentum entangled size.
If we were to scale up the crystal size, we would find entanglement between atoms far sooner than between nucleons.
(Leggett makes the same point in Ref.~\cite{Leggett2016Note} to draw an analogy with Cooper pairs in superconducting qubits.)
On the other hand, taking nucleons would actually increase the position entangled size of $\rho_{\SI{1}{s}}$ by a factor of about $12.5$, since the nucleon position uncertainty $\delta x$ is still dominated by each atom's centre-of-mass motion.
From now on, we choose to stop partitioning at the atomic scale, motivated by wanting the subsystems to be locally addressable, at least in principle.\\

\section{Oscillators} \label{sec:oscillator}

\subsection{Mode expansions and effective sizes} \label{sec:oscillator_basics}
Here, we apply the two measures to mechanical oscillators.
In particular, we demonstrate how to extend the scope of the measures to include arbitrary modes of oscillation.
We model such systems as collections of coupled oscillating particles labelled by their equilibrium positions $\vb*{r} \in \mathbb{R}^3$, with $\vb*{x}(\vb*{r})$ being the position observable (i.e., displacement from equilibrium) of the particle corresponding to $\vb*{r}$.
Here, we outline the results for position, with analogous momentum calculations in Appendix~\ref{app:oscillator}.
The modes of oscillation are indexed by wavenumbers $\vb*{k}$, with corresponding quadratures $\vb*{X}_{\vb*{k}}$.
These observables are related by a set of mode functions $w_{\vb*{k}}(\vb*{r})$ via
\begin{align} \label{eqn:mode_expansion}
    \vb*{x}(\vb*{r}) = \sum_{\vb*{k}} w_{\vb*{k}}(\vb*{r}) \vb*{X}_{\vb*{k}},
\end{align}
whose normalisations are the \emph{mode volumes} $V_{\vb*{k}}$:
\begin{align} \label{eqn:mode_orthogonality}
    \int \dd^3 \vb*{r} \, w_{\vb*{k}}(\vb*{r}) w_{\vb*{k'}}(\vb*{r}) = \delta_{\vb*{k},\vb*{k'}} V_{\vb*{k}}.
\end{align}
An extensive quantity can be formed by multiplying by the density (assumed constant $\varrho = M/V$):
\begin{align} \label{eqn:q_mode_expansion}
    \vb*{q}(\vb*{r}) & := \varrho \vb*{x}(\vb*{r}) = \sum_{\vb*{k}} \frac{w_{\vb*{k}} (\vb*{r})}{V_{\vb*{k}}} \vb*{Q}_{\vb*{k}}, \nonumber \\
    \vb*{Q}_{\vb*{k}} &:= M_{\vb*{k}} \vb*{X}_{\vb*{k}},
\end{align}
where $M_{\vb*{k}} = \varrho V_{\vb*{k}}$ is the \textbf{mode effective mass}.
Note that there is a freedom of choice in the normalisation of the mode functions: rescaling $w_{\vb*{k}} \to c w_{\vb*{k}}$ has the effect $V_{\vb*{k}} \to c^2 V_{\vb*{k}}$ and correspondingly $\vb*{Q}_{\vb*{k}} \to c \vb*{Q}_{\vb*{k}}$.
We choose the normalisation $\max_{\vb*{r}} \abs{w_{\vb*{k}}(\vb*{r})} = 1$, such that the magnitude of $\vb*{X}_{\vb*{k}}$ is the maximum oscillation amplitude of the mode.
This choice has the consequence $V_{\vb*{k}} \leq V$ and $M_{\vb*{k}} \leq M$ and it is natural to interpret these as the effective volume and mass occupied by the mode $\vb*{k}$.
For instance, the centre-of-mass mode has $w_\text{CM} = 1$, so $V_\text{CM} = V$, $M_\text{CM} = M$.

In experiments with quantum oscillators, there is typically a preferred mode of interest $\vb*{k}$, so we will calculate macroscopicity measures with respect to the observable $A = Q_{\vb*{k}}$, where from now on we focus on oscillation along a single direction $\vb*{\hat{z}}$, so $Q_{\vb*{k}} := \vb*{\hat{z}} \cdot \vb*{Q}_{\vb*{k}}$.
For the extensive size, we have
\begin{align}
    \extn(\rho, Q_{\vb*{k}}) = \frac{M_{\vb*{k}}^2 \mc{F}(\rho,X_{\vb*{k}})}{4Q_0^2}.
\end{align}

For the entangled size, we need to decompose this observable into its local terms $A = \int \dd^3 \vb*{r} \, A(\vb*{r})$, so use the orthogonality condition \eqref{eqn:mode_orthogonality} to obtain
\begin{align}
    A(\vb*{r}) = w_{\vb*{k}}(\vb*{r}) q(\vb*{r}).
\end{align}
We partition our system into disjoint spatial regions $R_i$ of equal volume $V / N$.
The local components of $A$ in each region are then
\begin{align} \label{eqn:mode_local_obs}
    A_i := \int_{R_i} \dd^3 \vb*{r} \, A(\vb*{r}) = \int_{R_i} \dd^3 \vb*{r} \, w_{\vb*{k}}(\vb*{r}) q(\vb*{r}) .
\end{align}

A thermal state of the system at temperature $T$ takes the form of a product over modes $\bigotimes_{\vb*{l}} \gamma_{\vb*{l}}$, where $\gamma_{\vb*{l}} \propto \sum_n e^{-n\hbar \omega_{\vb*{l}} / \kb T} \dyad{n}_{\vb*{l}}$ is the bosonic thermal state of mode $\vb*{l}$ having frequency $\omega_{\vb*{l}}$ and mean excitation number $\bar{n}_{\vb*{l}} = \left[ e^{\hbar \omega_{\vb*{l}} / \kb T} - 1 \right]^{-1}$.
In experiments that address a single mode $\vb*{k}$ and prepare it in a particular state $\rho_{\vb*{k}}$, we assume the whole system to be in the state\footnote{The assumption of thermality in all modes requires that there be no significant non-thermal sources of vibration from the environment, especially around the lower-lying frequencies which have the largest contributions to the single-particle vibrations.}
\begin{align} \label{eqn:addressed_state}
    \rho = \rho_{\vb*{k}} \ox \bigl( \bigotimes_{\vb*{l} \neq \vb*{k}} \gamma_{\vb*{l}} \bigr).
\end{align}
Given a state of the form \eqref{eqn:addressed_state}, we use Eq.~\eqref{eqn:mode_local_obs} to compute the sum of variances required for $\entn$ as
\begin{align} \label{eqn:modes_variance}
    \sum_i \vari (\rho, A_i) & = \varrho^2 \left[ \sum_i \zeta(i,\vb*{k},\vb*{k})^2 \right] \vari (\rho, X_{\vb*{k}}) + \varrho^2 \sum_{\vb*{l} \neq \vb*{k}} \left[ \sum_i \zeta(i,\vb*{k},\vb*{l})^2 \right] \nu_{\vb*{l}} (1 + 2\bar{n}_{\vb*{l}}), \\
    \text{where } \zeta(i,\vb*{k},\vb*{l}) & := \int_{R_i} \dd^3 \vb*{r} \, w_{\vb*{k}}(\vb*{r}) w_{\vb*{l}}(\vb*{r}),
\end{align}
and $\nu_{\vb*{l}} = \hbar / 2 M_{\vb*{l}} \omega_{\vb*{l}}$ is the ground-state position quadrature variance for mode $\vb*{l}$.
The coefficient $\zeta(i,\vb*{k},\vb*{l})$ measures the overlap between modes $\vb*{k}$ and $\vb*{l}$ within the region $R_i$.

We can ask how to choose the partition so as to maximise $\entn$.
Suppose we fine-grain each subsystem $\Pi_i$ into a sub-partition $\Pi_{i,j}$, with corresponding observables $A_i = \sum_j B_{i,j}$.
Since the variance can be expanded in terms of covariances as $\vari (\rho,A_i) = \sum_j \vari (\rho,B_{i,j}) + \sum_{k\neq j} \mathrm{Cov}(\rho,B_{i,j},B_{i,k})$, it is clear that fine-graining attains a smaller sum of variances, i.e., $\sum_{i,j} \vari (\rho,B_{i,j}) < \sum_i \vari (\rho,A_i)$, when all the covariances are positive.
We expect that nearby particles within a small volume comprising $\Pi_i$ should oscillate in phase -- and thus have positively correlated motions -- for any reasonable temperature in which the highest modes are not significantly excited, such that the size of this small volume is less than the wavelength of the lowest modes.
(This claim is verified in several works analysing vibrational correlations in different crystal structures~\cite{Steif1987Vibrational,Filipponi1988Vibrational,Tomaschitz2021Effective}.)
This fact suggests fine-graining as far as possible, to individual atoms.
In the continuum limit of regions $R_i$ whose sizes are much smaller than the scales over which both $w_{\vb*{k}}(\vb*{r})$ and $\vari (\rho, q(\vb*{r}))$ vary, we can approximate
\begin{align}
    A_i \approx U w_{\vb*{k}}(\vb*{r}_i) q(\vb*{r}_i),
\end{align}
where $R_i$ has volume $U = V/N$ and is centred at position $\vb*{r}_i$.
Then we have
\begin{align}
    \sum_i \vari (\rho, A_i) & \approx U \int \dd^3 \vb*{r} \, w_{\vb*{k}}(\vb*{r})^2\,  \vari (\rho, q(\vb*{r})).
\end{align}
    
For a thermal system, we approximate the position variance $\vari (\rho,x(\vb*{r}_i))$ of a single atom in the bulk of a three-dimensional object by a (temperature-dependent) constant $\Delta u^2$.
This quantity can in principle be calculated from Eq.~\eqref{eqn:modes_variance} using knowledge of the modes~\cite{Montroll1956Theory}, but is also measurable via the Debye-Waller factor in diffraction experiments~\cite{Filipponi1988Vibrational}.
In typical systems, as seen in the following sections, one has $\vari (\rho,X_{\vb*{k}}) \ll \Delta u^2$, since the zero-point position fluctuations are of the order $\SI{e-15}{m}$ while $\Delta u \sim \SI{e-11}{m}$.
However, since $\vari (\rho,q(\vb*{r}))$ has contributions from many modes, cooling down a single mode close to the ground state will not significantly affect it; in general, $\Delta u^2$ will if anything be a slight overestimate, giving a lower bound to the entangled size.
(If $\vb*{k}$ is the CM mode, then $\vari (\rho,X_{\vb*{k}})$ is not included in $\Delta u^2$, so contributes separately to the single-particle variance -- among the examples treated below, this applies only to the levitated nanoparticle of Ref.~\cite{Rossi2024Quantum}.)

The sum of variances when partitioned into single atoms thus simplifies to
\begin{align} \label{eqn:atoms_variance_sum}
    \sum_i \vari (\rho,A_i) & \approx U \int \dd^3 \vb*{r} \, w_{\vb*{k}}(\vb*{r})^2 \, \varrho^2 \Delta u^2 \nonumber \\
        & = \frac{V V_{\vb*{k}} \varrho^2 \Delta u^2}{N} \nonumber \\
        & = \frac{V M_{\vb*{k}}^2 \Delta u^2}{N V_{\vb*{k}}} .
\end{align}
The entangled size then becomes
\begin{align} \label{eqn:continuum_entn}
    \entn(\rho,Q_{\vb*{k}},\Pi) & \approx N \cdot \frac{V_{\vb*{k}}}{V} \cdot \frac{\mc{F}(\rho_{\vb*{k}}, X_{\vb*{k}})}{4 \Delta u^2} \nonumber \\
        & = N_{\vb*{k}} \cdot \frac{\mc{F}(\rho_{\vb*{k}}, X_{\vb*{k}})}{4 \Delta u^2}.
\end{align}
We see from Eq.~\eqref{eqn:continuum_entn} that $\entn$ is then the effective \textbf{mode particle number} $N_{\vb*{k}} = N V_{\vb*{k}}/V$, multiplied by the squared ratio of the mode's spatial quantum fluctuations to the single-atom position uncertainty.

\subsection{Estimates for thermal states} \label{sec:oscillator_simple}
In this section, we consider oscillators in which the mode of interest is in a thermal state at temperature $T$ with corresponding mean phonon number $\bar{n}$, which is often cooled to a lower temperature than the rest of the system.
We treat the chosen mode of interest as having effective mass $M_{\vb*{k}}$ and frequency $\omega_{\vb*{k}}$, and its QFI values are found to be (see Appendix~\ref{app:oscillator})
\begin{align}
    \frac{1}{4}\mc{F}(\rho, Q_{\vb*{k}}) &= \frac{\hbar M_{\vb*{k}}}{2 \omega_{\vb*{k}} (2\bar{n}+1)} = \frac{(M_{\vb*{k}} \xzp)^2}{2\bar{n}+1}, \nonumber \\
    \frac{1}{4}\mc{F}(\rho, P_{\vb*{k}}) & = \frac{\hbar M_{\vb*{k}} \omega_{\vb*{k}}}{2(2\bar{n}+1)} = \frac{(\hbar / 2 \xzp)^2}{2\bar{n}+1},
\end{align}
where $\xzp = \sqrt{\hbar / 2 M_{\vb*{k}} \omega_{\vb*{k}}}$ are the zero-point position fluctuations of mode $\vb*{k}$.
Thus, the extensive sizes are
\begin{align} \label{eqn:oscillator_extn}
    \extn(\rho, Q_{\vb*{k}}) & = \frac{\hbar M_{\vb*{k}}}{2 \omega_{\vb*{k}}(2\bar{n}+1) Q_0^2} = \frac{1}{2\bar{n}+1} \left(\frac{M_{\vb*{k}} \xzp}{m_u a_0}\right)^2, \nonumber \\
    \extn(\rho, P_{\vb*{k}}) & = \frac{\hbar M_{\vb*{k}} \omega_{\vb*{k}}}{2(2\bar{n}+1) P_0^2} = \frac{1}{2\bar{n}+1} \left(\frac{a_0}{\xzp}\right)^2.
\end{align}

From the expression in Eq.~\eqref{eqn:continuum_entn}, the entangled size for the single-atom partition is
\begin{align} \label{eqn:oscillator_entn}
    \entn(\rho, Q_{\vb*{k}},\Pi) \approx N_{\vb*{k}} \cdot \frac{1}{2\bar{n}+1} \cdot \left(\frac{\xzp}{\Delta u}\right)^2 .
\end{align}
(The momentum entangled size is not shown here as it turns out to be negligible.)
Hence we see the crucial dependence on temperature, going inversely with $\bar{n}$, and on the ratio of zero-point fluctuations of the mode to the atomic fluctuations.

In Table~\ref{tab:oscillators}, we show the results for a collection of mechanical oscillator experiments and proposals.
Refs.~\cite{Teufel2011Sideband,Verhagen2012Quantum} demonstrate cooling of the fundamental mode close to its ground state, while Ref.~\cite{Ringbauer2018Generation} uses a higher mass at room temperature.
In Ref.~\cite{Chegnizadeh2024Quantum}, six massive oscillators are cooled close to the ground state of a collective vibrational mode, which, relative to a single oscillator, enhances $\extn$ by a factor of 6, while $\entn$ is unchanged.
Ref.~\cite{Rossi2024Quantum} is a levitated nanoparticle prepared in a squeezed state, which differs from the other oscillators in that the addressed mode is the CM, and the single-particle position variance is dominated by the CM variance; so we replace $\Delta u$ by $\Delta X_\text{CM}$ in Eq.~\eqref{eqn:oscillator_entn}.
The proposal of Ref.~\cite{Pikovski2012Probing} is suggested to test quantum gravity models which predict modifications to the canonical commutation relations.
Ref.~\cite{Tobar2024Detecting} proposes signatures of gravitons via energy exchange with fundamental modes of very massive oscillators, which need to be cooled close to their ground state; we take two cases (the first and fifth columns of Table 1 of that paper) with different resonant frequencies.
For the proposals, we take $\Delta u = \SI{1e-11}{\meter}$, typical of solids at room temperature. Further details are given in Appendix~\ref{app:oscillator}.

\begin{table}
    \centering \scriptsize
    \addtolength{\leftskip} {-2cm} 
    \addtolength{\rightskip}{-2cm}
    \begin{tabular}{ l | l | l | l | l | l | l | l}
        Reference   & $M_{\vb*{k}}$ $(\si{\kilogram})$  & $\omega / 2\pi$ ($\si{\hertz}$)  &  $\Delta X_{\vb*{k}}^\mathrm{zp}$  ($\si{m}$) & $\bar{n}$    & $N_{\vb*{k}}$     & $\extn(Q_{\vb*{k}})$  & $\entn(Q_{\vb*{k}})$ \\ \hline
        Kienzler~(2016) \cite{Kienzler2016Observation}   & $\num{6.6e-26}$   & $\num{2.1e6}$     & $\num{7.8e-9}$    & n/a   & $1$ & $\num{1.5e9}$   & $1$ \\
        Teufel~(2011) \cite{Teufel2011Sideband}  & $\num{1.3e-14}$   & $\num{1.1e7}$     & $\num{7.8e-15}$   & $0.34$    & $\num{2.9e11}$    & $\num{7.9e17}$    & $\num{3.7e4}$ \\
        Verhagen~(2012) \cite{Verhagen2012Quantum}    &  $\num{3.2e-12}$  & $\num{7.8e7}$     & $\num{1.8e-16}$   & $1.7$ & $\num{9.8e13}$    & $\num{1.0e19}$    & $\num{1.2e3}$ \\
        Ringbauer~(2018) \cite{Ringbauer2018Generation}   & $\num{1.1e-10}$   & $\num{1.1e5}$ & $\num{8.3e-16}$   & $\num{6.0e7}$   & $\num{3.5e15}$    & $\num{9.8e15}$    & $\num{5.0e-2}$ \\
        Chegnizadeh~(2024) \cite{Chegnizadeh2024Quantum}  & $\num{1.3e-11}$   & $\num{2.0e6}$ & $\num{5.6e-16}$   & $0.4$   & $\num{3.0e14}$  & $\num{4.1e21}$    & $\num{1.8e5}$ \\
        Bild~(2023)~\cite{Bild2023Schrodinger}    & $\num{4.0e-9}$ & $\num{5.0e9}$     & $\num{6.5e-19}$   & n/a  & $\num{1.2e17}$    & $\num{3.0e21}$    & $\num{4.1e3}$ \\ 
        Rossi~(2024)~\cite{Rossi2024Quantum}  & $\num{1.2e-18}$   & $\num{5.7e4}$    & $\num{1.1e-11}$   & n/a   & $\num{3.6e7}$ & $\num{9.9e17}$    & $\num{1.3e7}$ \\ \hline
        Pikovski~(2012) \cite{Pikovski2012Probing}   & $\num{1.0e-11}$ & $\num{1.0e5}$     & $\num{2.9e-15}$   & $\num{30}$    & $\num{3.0e14}$ &  $\num{1.8e21}$  & $\num{4.1e5}$ \\
        Tobar~(2024) \cite{Tobar2024Detecting} (a)   & $7.5$  & $\num{1.0e2}$ & $\num{1.1e-19}$   & $\approx 0$   & $\num{5.0e26}$   & $\num{8.2e37}$  &$\num{5.6e10}$ \\
        Tobar~(2024) \cite{Tobar2024Detecting} (b) & $\num{2.6e4}$     & $\num{1.1e3}$     & $\num{5.5e-22}$   & $\approx 0$   & $\num{1.7e29}$    & $\num{2.6e40}$    & $\num{5.1e8}$
    \end{tabular}
    \caption{Experimental parameters and estimated extensive and entangled sizes for mechanical oscillator experiments (upper set of rows) and proposals (lower set of rows). For the Bild et al.\ experiment~\cite{Bild2023Schrodinger}, we take the non-Gaussian state displayed in Fig.~\ref{fig:ExpWigners}(c). The effective mass for Ref.~\cite{Bild2023Schrodinger} differs from the value quoted in that reference due to our normalisation $\max_{\vb*{r}} \abs{w_{\vb*{k}}(\vb*{r})} = 1$.}
    \label{tab:oscillators}
\end{table}

\subsection{Estimates from Wigner function}
Here, we apply our measures to cases in which the QFI is more directly known -- this enables an extension beyond thermal states.
Experiments with massive oscillators allow for estimating the QFI in different ways.
For example, in optomechanical and electromechanical setups we are able to use the (optical or microwave) electromagnetic field interacting with the oscillating mass to measure its phase space quadratures.
From this, a lower bound to the QFI is given by a measurement of the oscillator's quantum noise level compared to the ground state noise (i.e., squeezing).

Alternatively, platforms where an oscillator is coupled to a two-level system, such as in trapped ions where the external motional state is coupled to the internal state, or in mechanical resonators coupled to superconducting qubits, have shown the capability of performing full state tomography of the oscillator's state. This is commonly achieved by taking displaced parity measurements of the phonon number, which correspond to direct measurement of the Wigner function value at a point in phase space. By scanning through many points it is possible to take 1-dimensional slices, or even full 2-dimensional information, of the Wigner function, which directly allows for computing the QFI \cite{Frowis2017Lower,Marti2024Quantum}. 

\begin{figure}[h]
    \centering
    \includegraphics[width=.9\textwidth]{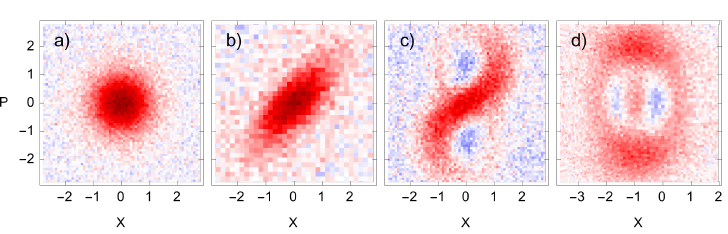}
    \caption{Wigner functions of quantum states prepared in a solid-state mechanical oscillator: a) Ground state, b) squeezed state, c) non-Gaussian state, and d) cat state. The corresponding QFI values for displacements are $\mc{F}(\rho,\hat{A})=$ $2.0$, $4.2$, $7.0$, $3.9$, respectively, where $\hat{A}$ is the dimensionless phase-space quadrature maximising the QFI (normalised such that coherent states give $\mc{F}=2$). Experimental data from \cite{Bild2023Schrodinger,Marti2024Quantum}.}
    \label{fig:ExpWigners}
\end{figure}

Here, we use Wigner function data from Refs.~\cite{Bild2023Schrodinger,Marti2024Quantum}.
The system is a high-order overtone mode (around the $500$th harmonic) of a sapphire ($\mathrm{Al}_2 \mathrm{O}_3$) crystal, density $\varrho = \SI{3.98e3}{\kilogram \meter^{-3}}$, with length $\SI{435}{\micro \meter}$ and a Gaussian profile of waist $\SI{27}{\micro \meter}$.
The precise values of $\Delta u$ are known from experimental data at room temperature~\cite[Table 1]{French1994Interband} -- these are almost the same for both types of atom, at $\SI{6.5e-12}{m}$.
In Fig.~\ref{fig:ExpWigners}, we show Wigner function data for four states: a) the ground state, b) a squeezed state, c) a more complex non-Gaussian state, and d) a cat state.
The QFI is computed with respect to the dimensionless quadratures $\hat X_{\vb*{k}} = X_{\vb*{k}} / (\sqrt{2}\Delta X_{\vb*{k}}^\mathrm{zp})$, $\hat P_{\vb*{k}} = P_{\vb*{k}} / (\sqrt{2}\Delta P_{\vb*{k}}^\mathrm{zp})$, with $\Delta P_{\vb*{k}}^\mathrm{zp} = \hbar/2\Delta X_{\vb*{k}}^\mathrm{zp}$, normalised as in Ref.~\cite{Marti2024Quantum} such that the ground state gives $\mc{F}=2$, and is maximised over phase-space rotations.
The resulting extensive and entangled sizes are shown in Table~\ref{tab:oscillators}; the largest values are found when the QFI-maximising direction is the position quadrature. \\

One-dimensional slices of the Wigner function are used in Ref.~\cite{Frowis2017Lower} to estimate the QFI in an experiment~\cite{Kienzler2016Observation} with a ${}^{40}\mathrm{Ca}^+$ ion in a harmonic trap of frequency $\omega = 2\pi \times \SI{2.08}{MHz}$.
This is placed in a cat state (superposition of coherent states) $\ket{\alpha} + \ket{-\alpha}$ with $\alpha = 5.9$, corresponding to a relative displacement of around $\SI{180}{nm}$~\cite{Kienzler2016Observation}.
Using the QFI estimated in Ref.~\cite{Frowis2017Lower}, we obtain $\extn(\rho,Q) \approx \num{1.5e9}$ (an ideal pure cat state with the same $\alpha$ would give $\approx \num{4.9e9}$).
There is of course only a single ion, so the entangled size is irrelevant.
(Even partitioning into nucleus plus electrons [so $n = 1 + 19$] leads to only $\entn(\rho,Q,\Pi) \approx 0.16$, as a result of the negligible contribution of the electrons due to their small mass.)

Note that Ref.~\cite{Frowis2017Lower} defines an effective size measure based on the QFI with a different normalisation, instead requiring us to identify a set $\mathcal{C}$ of ``classical'' states.
The denominator in Eq.~\eqref{eqn:entn_def} is correspondingly replaced by $\max_{\psi_\text{cl} \in \mathcal{C}} \mc{F}(\psi_\text{cl}, A)$.
If the classical states are product states with respect to the partition $\Pi$, then a sum of variances is recovered -- however, not necessarily involving the state $\rho$ under consideration.
While a set of classical states may be reasonably chosen in simple cases like sets of qubits or single bosonic modes, there is ambiguity more generally.
For example, there is no intrinsic classical length or momentum scale for a free particle coherent state.
For the experiment of Ref.~\cite{Kienzler2016Observation}, coherent states of the oscillator are chosen as the classical reference, giving an effective size of roughly $43$.

\section{Diffraction} \label{sec:diffraction}

Experiments on the diffraction of large molecules operate by collimating a molecular beam, selecting a range of velocities, passing them through a grating, and finally detecting them at some plane further down their flight path~\cite{Hornberger2012Colloquium,Nimmrichter2014Macroscopic}.
Over time, detection events build up an interference pattern.
Such a pattern is an indication that the molecule was coherently delocalised over a significant length of the grating, its wavepacket spanning multiple slits.
Therefore we are interested in the extensive and entangled sizes of the state at the grating, using the transverse position variable $Q$.

We show here that the relevant QFI can be lower-bounded from the detection statistics.
This works in principle for any grating--detector distance -- we assume only that the particles evolve freely after the grating.
We need to estimate $\mc{F}(\rho_0,Q)$ from statistics collected after a time of flight $t$, from measurements on the state $\rho_t$.
Let $p(x)$ be the probability density that a particle is detected at transverse position $x$ (assuming a perfectly efficient detector).
Then we have the inequality
\begin{align} \label{eqn:diffraction_qfi_bound}
    \mc{F}(\rho_0, Q) \geq (\hbar t)^2 \mc{F}_\text{cl}[p(x)] = (\hbar t)^2 \int \dd x \, \frac{p'(x)^2}{p(x)},
\end{align}
where the quantity on the right-hand side is the classical FI of the distribution $p(x)$ under displacements.

To prove Eq.~\eqref{eqn:diffraction_qfi_bound}, let $U_t$ denote the free time-evolution unitary.
The transverse position $X$ and momentum $P$ evolve in the Heisenberg picture as $U_t^\dagger X U_t = X + \frac{t}{M}P$, $U_t^\dagger P U_t = P$.
For the QFI of the state at time $t$ under the generator $X - \frac{t}{M}P$, we have
\begin{align}
    \mc{F}\left(\rho_t, X - \frac{t}{M}P \right) & = \mc{F}\left( U_t^\dagger \rho_t U_t, U_t^\dagger \left[ X - \frac{t}{M} P \right] U_t \right) \nonumber \\
        & = \mc{F}(\rho_0, X).
\end{align}
The effect of the generator $X - \frac{t}{M}P$ at the detector is to shift the position-space wavefunction in the transverse direction and add a position-dependent phase.
Since the measurement only depends on the probability density $p(x) = \ev{\rho_t}{x}$, this phase is not observed.
The generator of spatial translations is $P/\hbar$; since the classical FI never exceeds the QFI, we have
\begin{align}
    \mc{F}(\rho_0, X) \geq \left( \frac{\hbar t}{M} \right)^2 \mc{F}_\text{cl}[p(x)] .
\end{align}
Rearranging gives Eq.~\eqref{eqn:diffraction_qfi_bound}.\\

In Appendix~\ref{app:diffraction}, we show how this bound can be applied to the Talbot-Lau interferometer setup from Ref.~\cite{Fein2019Quantum}.
This experiment uses a range of molecules of $N \sim \num{2000}$ atoms, with most common mass $M = \num{26777} m_u$.
They diffract at a central grating laser of period $\SI{266}{nm}$ and are detected after a further distance of $\SI{1}{m}$.
The final distribution $p(x)$ is not directly measured -- rather, the number of molecules passing through a final grating that is scanned transversely -- but we show that the classical FI is still possible to lower-bound from the observed statistics.
We obtain an extensive size $\extn(\rho_0,Q) \gtrsim \num{1.4e14}$.
The entangled size requires an additional estimate of the centre-of-mass position uncertainty, and results in a witness of entanglement between $\entn(\rho_0,Q,\Pi) \sim 5$ atoms.

We also apply this bound to the recent Talbot-Lau experiment with sodium nanoparticles of Ref.~\cite{Pedalino2026Probing}, with mean mass $M = \num{1.72e5} m_u$ ($N \approx \num{7.5e3}$ atoms).
We obtain $\extn(\rho_0,Q) \gtrsim \num{2e13}$ and $\entn(\rho_0,Q,\Pi) \sim 80$ (see Appendix~\ref{app:diffraction}).
Note that the bound~\eqref{eqn:diffraction_qfi_bound} is independent of the particle mass; here, it is limited by the lower fringe visibility and grating transmission compared with Ref.~\cite{Fein2019Quantum}.

\section{Conclusions and outlook}
\label{sec:conclusions}

In summary, we have introduced two measures of quantum macroscopicity that capture different ways a system can display large-scale quantum behaviour: an extensive size measuring the amount of quantum coherence in atomic-scale units, and an entangled size quantifying the number of entangled particles.
Both require a choice of relevant observable -- here, we use centre-of-mass position scaled by total mass, or momentum -- and the latter requires a sensible partition into particles, here typically single atoms.

\begin{figure}[h]
    \centering
    \includegraphics[width=0.95\textwidth]{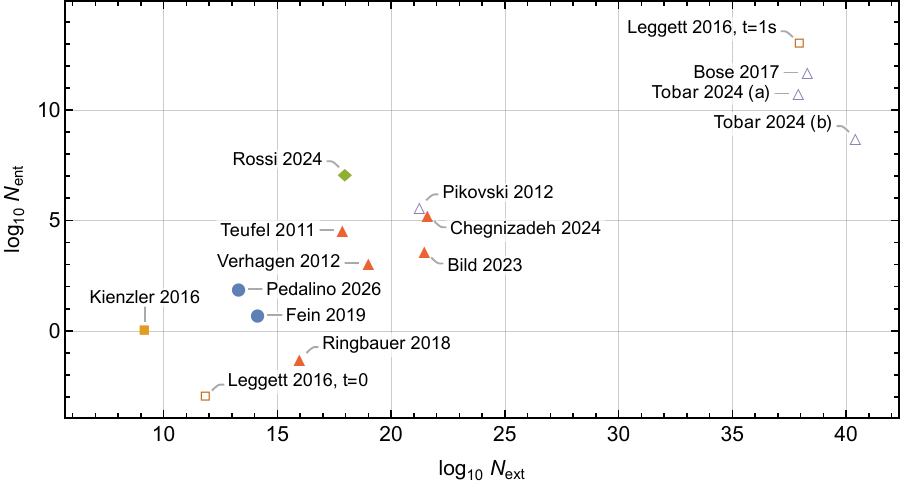}
    \caption{Extensive and entangled sizes for all systems considered in this work. Each value is calculated with respect to the position degree of freedom (except for the point labelled Leggett 2016, $t=0$), and the entangled size is for a partition into single atoms. Experiments: mechanical oscillators (filled triangles), levitated nanoparticle (filled diamond), trapped ion (filled square), diffracting molecules and nanoparticles (filled circle). Hypothetical systems: proposed experiments (unfilled triangles), Leggett's crystal (unfilled squares).}
    \label{fig:plot}
\end{figure}

Fig.~\ref{fig:plot} plots the size values for all cases we have considered, comprising the experimental estimates, as well as Leggett's thought experiment and proposals.
Although the two size measures are in principle independent, we see that they are quite well correlated in realistic cases.
The more recent experiments in optomechanics and levitated nanoparticles are leading in both measures.
Note however that our estimates are somewhat more stringent for Bild et al.~\cite{Bild2023Schrodinger} and for the diffracted molecules in Fein et al.~\cite{Fein2019Quantum}, relying on lower bounds to the QFI directly from measurement data, which may not be tight.
Furthermore, the nanoparticle centre-of-mass coherence length reported in Ref.~\cite{Rossi2024Quantum} is obtained from retrodiction using continuous-time measurement data and so depends on additional assumptions about the system's dynamics.
It is also interesting that entanglement depth comparable to the largest numbers found here has been demonstrated in collective spins of atomic ensembles~\cite{Frowis2017Experimental}.
There is a notable separation between current experiments and the graviton-detection proposals of Ref.~\cite{Tobar2024Detecting} of up to roughly 20 orders of magnitude in $\extn$ and 5 in $\entn$, while the proposal of Ref.~\cite{Pikovski2012Probing} is comparable to current experiments in both measures.
For completeness, we also include the proposal of Ref.~\cite{Bose2017Spin} to test the quantumness of gravity via its potential to entangle two separated massive objects (using a mass $M \approx \SI{e-14}{\kilogram}$ and separation $\Delta X_\text{CM} \approx \SI{250}{\micro \meter}$).
It is also noteworthy that these quantum gravity proposals are of very similar magnitude in both measures, despite their aim to test different aspects of quantisation.
What this implies about their feasibility is of course open to debate.

We now comment on possible extensions.
Instead of taking only $Q$ and $P$, one could use linear combinations $R(t) = Q + tP$, varying over all phase-space directions, with the associated atomic-scale unit $R_0(t)^2 := Q_0^2 + t^2 P_0^2$.
This way, $\extn(\rho, R(t))$ can be maximised over all $t$ and includes $\extn(\rho,Q)$ and $\extn(\rho,P)$ as special cases in the limits $t \to 0$ and $t \to \infty$, respectively.
Angular momentum is another possible mechanical observable to consider; the atomic-scale unit is simply $J_0 = \frac{Q_0}{m_u} \cdot P_0 = \frac{\hbar}{2}$.
It is interesting to note that $R(t)$ appears in the time-space component of the relativistic angular momentum tensor constructed from the position and momentum four-vectors, hinting at a potential relativistic generalisation.

Beyond mechanical systems, in particular the macroscopicity of superconducting qubits \cite{Clarke2008Superconducting} has been long debated \cite{Leggett1980Macroscopic,Bjork2004Size,Leggett2016Note,Korsbakken2009Electronic}.
The earliest strong experimental evidence for superpositions of different current states in flux qubits was obtained by spectroscopic methods \cite{vanderWal2000Quantum,Friedman2000Quantum}.
Such states may also be considered as mechanical superpositions by converting a current $I$ to a total momentum integrated along the circuit, $P = m_e l I/e$, where $m_e$ and $e$ are the mass and charge of the electron and $l$ is the length of the circuit. Assuming a pure state, the extensive size in momentum may be estimated as $\extn(\rho,P) = ( m_e l \Delta I / 2 e P_0 )^2$, where $\Delta I$ is the difference in current between the branches.
From Ref.~\cite{Friedman2000Quantum}, $\Delta I \approx \SI{2}{\micro \ampere},\, l = \SI{560}{\micro \meter}$, thus $\extn(\rho,P) \approx \num{1e7}$ -- not yet approaching the value of $\num{7e11}$ in Leggett's original crystal example.
For the entangled size, the relevant partition according to Leggett~\cite{Leggett2016Note} is into Cooper pairs, as opposed to individual electrons as performed in other analyses~\cite{Korsbakken2009Electronic,Korsbakken2010Size}.
The zero-current state (the Bardeen-Cooper-Schrieffer ground state) can be regarded approximately as a condensate of $N_\text{pairs}$ pairs in a zero-momentum eigenstate~\cite{IbachLuth}.
The momentum shift between branches per pair is $\Delta p = \Delta P / N_\text{pairs} \approx \SI{6e-30}{\kilogram \meter \second^{-1}}$, corresponding to a length scale $\hbar / \Delta p \sim \SI{20}{\micro \meter}$.
The coherence length of the Cooper pair centre of mass must be at least the size of the circuit, thus $\delta p \ll \Delta p$, and we are in the ``full cat'' regime where $\entn(\rho,P,\Pi_\text{pairs}) \sim N_\text{pairs} \sim \num{e9}$.
This agrees with Leggett's conclusion.
It should, however, be noted that the evidence for coherence in these early experiments is also consistent with significantly lower QFI than naively estimated here~\cite{Frowis2018Insufficiency}.

Other systems we leave for future work are Bose-Einstein condensates and quantum optics.
For the latter, the observable $Q$ must be redefined since photons do not have mass -- one possibility is $\int \dd^3 \vb*{r}\, x \mc{E}(\vb*{r}) / c^2$, where $\mc{E}$ is the energy density of the electromagnetic field.
The ability to extend to optical systems may be an advantage over approaches relying on massive particles, such as those based on minimal modifications to quantum dynamics.
For systems of identical particles generally, one has to be careful in interpreting entanglement, but this can be put on a secure footing~\cite{Morris2020Entanglement}.
We hope that our work can thus lead to a unifying formalism for evaluating the quantum macroscopicity of a wide range of experimental systems in a transparent and comparable way.

\begin{acknowledgments}
    We gratefully acknowledge Vlatko Vedral, Stephen Barnett, and Stefan Nimmrichter for discussions, as well as Myungshik Kim for feedback on an early draft.
    This project has received funding from the European Union's Horizon 2020 research and innovation programme under the Marie Sk\l odowska-Curie grant agreement No.~945422.
    This work has been supported by the Deutsche Forschungsgemeinschaft (DFG, German Research Foundation, project numbers 447948357 and 440958198), the Sino-German Center for Research Promotion (Project M-0294), and the German Ministry of Education and Research (Project QuKuK, BMBF Grant No.~16KIS1618K).
    MF was supported by the Swiss National Science Foundation Ambizione Grant No. 208886, and The Branco Weiss Fellowship -- Society in Science, administered by the ETH Z\"{u}rich.
\end{acknowledgments}

\vspace{4mm}
\textbf{Author contributions.–} B.Y. proposed the original idea and developed the theoretical formalism. M.F. contributed to the computation of concrete examples and provided experimental considerations. Both authors checked the results presented and contributed to the writing and revision of the manuscript.

\bibliographystyle{quantum}
\bibliography{refs2}

\begin{thebibliography}{10}

\bibitem{Schrodinger1935Die}
E.~Schr{\"o}dinger.
\newblock ``Die gegenw{\"a}rtige {Situation} in der {Quantenmechanik}''.
\newblock \href{https://dx.doi.org/10.1007/BF01491891}{Die Naturwissenschaften
  {\bf 23}, 807--812}~(1935).

\bibitem{Zurek2003Decoherence}
Wojciech~Hubert Zurek.
\newblock ``Decoherence, einselection, and the quantum origins of the
  classical''.
\newblock \href{https://dx.doi.org/10.1103/RevModPhys.75.715}{Reviews of Modern
  Physics {\bf 75}, 715--775}~(2003).

\bibitem{Peres1964Macroscopic}
Asher Peres and Nathan Rosen.
\newblock ``Macroscopic bodies in quantum theory''.
\newblock \href{https://dx.doi.org/10.1103/PhysRev.135.B1486}{Physical Review
  {\bf 135}, B1486}~(1964).

\bibitem{Kofler2007Classical}
Johannes Kofler and {\v{C}}aslav Brukner.
\newblock ``Classical world arising out of quantum physics under the
  restriction of coarse-grained measurements''.
\newblock \href{https://dx.doi.org/10.1103/PhysRevLett.99.180403}{Physical
  Review Letters {\bf 99}, 180403}~(2007).

\bibitem{Raeisi2011Coarse}
Sadegh Raeisi, Pavel Sekatski, and Christoph Simon.
\newblock ``Coarse graining makes it hard to see micro-macro entanglement''.
\newblock \href{https://dx.doi.org/10.1103/PhysRevLett.107.250401}{Physical
  Review Letters {\bf 107}, 250401}~(2011).

\bibitem{Sekatski2014Size}
Pavel Sekatski, Nicolas Sangouard, and Nicolas Gisin.
\newblock ``Size of quantum superpositions as measured with classical
  detectors''.
\newblock \href{https://dx.doi.org/10.1103/PhysRevA.89.012116}{Physical Review
  A {\bf 89}, 012116}~(2014).

\bibitem{Skotiniotis2017Macroscopic}
Michalis Skotiniotis, Wolfgang D{\"u}r, and Pavel Sekatski.
\newblock ``Macroscopic superpositions require tremendous measurement
  devices''.
\newblock \href{https://dx.doi.org/10.22331/q-2017-11-21-34}{Quantum {\bf 1},
  34}~(2017).

\bibitem{Frowis2018Macroscopic}
Florian Fr{\"o}wis, Pavel Sekatski, Wolfgang D{\"u}r, Nicolas Gisin, and
  Nicolas Sangouard.
\newblock ``Macroscopic quantum states: Measures, fragility, and
  implementations''.
\newblock \href{https://dx.doi.org/10.1103/RevModPhys.90.025004}{Reviews of
  Modern Physics {\bf 90}, 025004}~(2018).

\bibitem{Shimizu2005Detection}
Akira Shimizu and Tomoyuki Morimae.
\newblock ``Detection of macroscopic entanglement by correlation of local
  observables''.
\newblock \href{https://dx.doi.org/10.1103/PhysRevLett.95.090401}{Physical
  Review Letters {\bf 95}, 090401}~(2005).

\bibitem{Lee2011Quantification}
Chang~Woo Lee and Hyunseok Jeong.
\newblock ``Quantification of macroscopic quantum superpositions within phase
  space''.
\newblock \href{https://dx.doi.org/10.1103/PhysRevLett.106.220401}{Physical
  Review Letters {\bf 106}, 220401}~(2011).

\bibitem{Cavalcanti2008Criteria}
E.~G. Cavalcanti and M.~D. Reid.
\newblock ``Criteria for generalized macroscopic and mesoscopic quantum
  coherence''.
\newblock \href{https://dx.doi.org/10.1103/PhysRevA.77.062108}{Physical Review
  A {\bf 77}, 062108}~(2008).

\bibitem{Yadin2016General}
Benjamin Yadin and Vlatko Vedral.
\newblock ``General framework for quantum macroscopicity in terms of
  coherence''.
\newblock \href{https://dx.doi.org/10.1103/PhysRevA.93.022122}{Physical Review
  A {\bf 93}, 022122}~(2016).

\bibitem{Dur2002Effective}
Wolfgang D{\"u}r, Christoph Simon, and J.~Ignacio Cirac.
\newblock ``Effective size of certain macroscopic quantum superpositions''.
\newblock \href{https://dx.doi.org/10.1103/PhysRevLett.89.210402}{Physical
  Review Letters {\bf 89}, 210402}~(2002).

\bibitem{Yadin2015Quantum}
Benjamin Yadin and Vlatko Vedral.
\newblock ``Quantum macroscopicity versus distillation of macroscopic
  superpositions''.
\newblock \href{https://dx.doi.org/10.1103/PhysRevA.92.022356}{Physical Review
  A {\bf 92}, 022356}~(2015).

\bibitem{Korsbakken2007Measurement}
Jan~I. Korsbakken, K.~Birgitta Whaley, Jonathan Dubois, and J.~Ignacio Cirac.
\newblock ``Measurement-based measure of the size of macroscopic quantum
  superpositions''.
\newblock \href{https://dx.doi.org/10.1103/PhysRevA.75.042106}{Physical Review
  A {\bf 75}, 042106}~(2007).

\bibitem{Marquardt2008Measuring}
Florian Marquardt, Benjamin Abel, and Jan von Delft.
\newblock ``Measuring the size of a quantum superposition of many-body
  states''.
\newblock \href{https://dx.doi.org/10.1103/PhysRevA.78.012109}{Physical Review
  A {\bf 78}, 012109}~(2008).

\bibitem{Nimmrichter2013Macroscopicity}
Stefan Nimmrichter and Klaus Hornberger.
\newblock ``Macroscopicity of mechanical quantum superposition states''.
\newblock \href{https://dx.doi.org/10.1103/PhysRevLett.110.160403}{Physical
  Review Letters {\bf 110}, 160403}~(2013).

\bibitem{Schrinski2019Macroscopicity}
Bj{\"o}rn Schrinski, Stefan Nimmrichter, Benjamin~A. Stickler, and Klaus
  Hornberger.
\newblock ``Macroscopicity of quantum mechanical superposition tests via
  hypothesis falsification''.
\newblock \href{https://dx.doi.org/10.1103/PhysRevA.100.032111}{Physical Review
  A {\bf 100}, 032111}~(2019).

\bibitem{Schrinski2023Macroscopic}
Bj{\"o}rn Schrinski, Yu~Yang, Uwe von L{\"u}pke, Marius Bild, Yiwen Chu, Klaus
  Hornberger, Stefan Nimmrichter, and Matteo Fadel.
\newblock ``Macroscopic quantum test with bulk acoustic wave resonators''.
\newblock \href{https://dx.doi.org/10.1103/PhysRevLett.130.133604}{Physical
  Review Letters {\bf 130}, 133604}~(2023).

\bibitem{Leggett1980Macroscopic}
Anthony~J. Leggett.
\newblock ``Macroscopic quantum systems and the quantum theory of
  measurement''.
\newblock \href{https://dx.doi.org/10.1143/PTPS.69.80}{Progress of Theoretical
  Physics Supplement {\bf 69}, 80--100}~(1980).

\bibitem{Frowis2012Measures}
Florian Fr{\"o}wis and Wolfgang D{\"u}r.
\newblock ``Measures of macroscopicity for quantum spin systems''.
\newblock \href{https://dx.doi.org/10.1088/1367-2630/14/9/093039}{New Journal
  of Physics {\bf 14}, 093039}~(2012).

\bibitem{Oudot2015Two}
E.~Oudot, P.~Sekatski, F.~Fr{\"o}wis, N.~Gisin, and N.~Sangouard.
\newblock ``Two-mode squeezed states as {Schr{\"o}dinger} cat-like states''.
\newblock \href{https://dx.doi.org/10.1364/JOSAB.32.002190}{Journal of the
  Optical Society of America B {\bf 32}, 2190--2197}~(2015).

\bibitem{Leggett2002Testing}
Anthony~J. Leggett.
\newblock ``Testing the limits of quantum mechanics: motivation, state of play,
  prospects''.
\newblock \href{https://dx.doi.org/10.1088/0953-8984/14/15/201}{Journal of
  Physics: Condensed Matter {\bf 14}, R415--R451}~(2002).

\bibitem{Braunstein1994Statistical}
Samuel~L. Braunstein and Carlton~M. Caves.
\newblock ``Statistical distance and the geometry of quantum states''.
\newblock \href{https://dx.doi.org/10.1103/PhysRevLett.72.3439}{Physical Review
  Letters {\bf 72}, 3439--3443}~(1994).

\bibitem{Shimizu2002Stability}
Akira Shimizu and Takayuki Miyadera.
\newblock ``Stability of quantum states of finite macroscopic systems against
  classical noises, perturbations from environments, and local measurements''.
\newblock \href{https://dx.doi.org/10.1103/PhysRevLett.89.270403}{Physical
  Review Letters {\bf 89}, 270403}~(2002).

\bibitem{Frowis2015Linking}
F.~Fr{\"o}wis, N.~Sangouard, and N.~Gisin.
\newblock ``Linking measures for macroscopic quantum states via photon--spin
  mapping''.
\newblock \href{https://dx.doi.org/10.1016/j.optcom.2014.07.017}{Optics
  Communications {\bf 337}, 2--11}~(2015).

\bibitem{Marvian2022Operational}
Iman Marvian.
\newblock ``Operational interpretation of quantum {Fisher} information in
  quantum thermodynamics''.
\newblock \href{https://dx.doi.org/10.1103/PhysRevLett.129.190502}{Physical
  Review Letters {\bf 129}, 190502}~(2022).

\bibitem{Yadin2017Thesis}
Benjamin Yadin.
\newblock ``Resource theories of quantum coherence: foundations and
  applications''.
\newblock \href{https://dx.doi.org/10.5287/ora-wx05gxk92}{{DPhil} thesis}.
\newblock University of Oxford.
\newblock ~(2017).

\bibitem{Nimmrichter2014Macroscopic}
Stefan Nimmrichter.
\newblock ``Macroscopic matter wave interferometry''.
\newblock \href{https://dx.doi.org/10.1007/978-3-319-07097-1}{Springer Theses}.
  Springer International Publishing. Cham~(2014).

\bibitem{Fadel2023Multiparameter}
Matteo Fadel, Benjamin Yadin, Yuping Mao, Tim Byrnes, and Manuel Gessner.
\newblock ``Multiparameter quantum metrology and mode entanglement with
  spatially split nonclassical spin ensembles''.
\newblock \href{https://dx.doi.org/10.1088/1367-2630/ace1a0}{New Journal of
  Physics {\bf 25}, 073006}~(2023).

\bibitem{Guhne2005Multipartite}
Otfried G{\"u}hne, G{\'e}za T{\'o}th, and Hans~J. Briegel.
\newblock ``Multipartite entanglement in spin chains''.
\newblock \href{https://dx.doi.org/10.1088/1367-2630/7/1/229}{New Journal of
  Physics {\bf 7}, 229}~(2005).

\bibitem{Leggett2016Note}
Anthony~J. Leggett.
\newblock ``Note on the ``size'' of {Schroedinger} cats''~(2016).
\newblock  \href{http://arxiv.org/abs/1603.03992}{arXiv:1603.03992}.

\bibitem{Cartz1955Thermal}
L.~Cartz.
\newblock ``Thermal vibrations of atoms in cubic crystals {II}: The amplitude
  of atomic vibrations''.
\newblock \href{https://dx.doi.org/10.1088/0370-1301/68/11/321}{Proceedings of
  the Physical Society. Section B {\bf 68}, 957--967}~(1955).

\bibitem{Bjork2004Size}
G.~Bj{\"o}rk and P.~G.~L. Mana.
\newblock ``A size criterion for macroscopic superposition states''.
\newblock \href{https://dx.doi.org/10.1088/1464-4266/6/11/001}{Journal of
  Optics B: Quantum and Semiclassical Optics {\bf 6}, 429--436}~(2004).

\bibitem{Steif1987Vibrational}
A.~Steif, S.~C. Tiersten, and S.~C. Ying.
\newblock ``Vibrational correlation functions for {Si} and {Ge}''.
\newblock \href{https://dx.doi.org/10.1103/PhysRevB.35.857}{Physical Review B
  {\bf 35}, 857--859}~(1987).

\bibitem{Filipponi1988Vibrational}
Adriano Filipponi.
\newblock ``Vibrational correlation function in ordered and disordered covalent
  solids''.
\newblock \href{https://dx.doi.org/10.1103/PhysRevB.37.7027}{Physical Review B
  {\bf 37}, 7027--7037}~(1988).

\bibitem{Tomaschitz2021Effective}
Roman Tomaschitz.
\newblock ``Effective real-space correlations of crystal lattice vibrations''.
\newblock \href{https://dx.doi.org/10.1016/j.jpcs.2020.109773}{Journal of
  Physics and Chemistry of Solids {\bf 152}, 109773}~(2021).

\bibitem{Montroll1956Theory}
Elliott~W. Montroll.
\newblock ``Theory of the vibration of simple cubic lattices with nearest
  neighbor interactions''.
\newblock In Jerzy Neyman, editor, Proceedings of the Third Berkeley Symposium
  on Mathematical Statistics and Probability, Volume 3: Contributions to
  Astronomy and Physics.
\newblock \href{https://dx.doi.org/10.1525/9780520350694-014}{Volume~3, pages
  209--246}.
\newblock Berkeley~(1956). University of California Press.

\bibitem{Rossi2024Quantum}
Massimiliano Rossi, Andrei Militaru, Nicola Carlon~Zambon, Andreu
  Riera-Campeny, Oriol Romero-Isart, Martin Frimmer, and Lukas Novotny.
\newblock ``Quantum delocalization of a levitated nanoparticle''.
\newblock \href{https://dx.doi.org/10.1103/2yzc-fsm3}{Physical Review Letters
  {\bf 135}, 083601}~(2025).
\newblock  \href{http://arxiv.org/abs/2408.01264}{arXiv:2408.01264}.

\bibitem{Teufel2011Sideband}
J.~D. Teufel, T.~Donner, D.~Li, J.~W. Harlow, M.~S. Allman, K.~Cicak, A.~J.
  Sirois, J.~D. Whittaker, K.~W. Lehnert, and R.~W. Simmonds.
\newblock ``Sideband cooling of micromechanical motion to the quantum ground
  state''.
\newblock \href{https://dx.doi.org/10.1038/nature10261}{Nature {\bf 475},
  359--363}~(2011).

\bibitem{Verhagen2012Quantum}
E.~Verhagen, S.~Del{\'e}glise, S.~Weis, A.~Schliesser, and T.~J. Kippenberg.
\newblock ``Quantum-coherent coupling of a mechanical oscillator to an optical
  cavity mode''.
\newblock \href{https://dx.doi.org/10.1038/nature10787}{Nature {\bf 482},
  63--67}~(2012).

\bibitem{Ringbauer2018Generation}
M.~Ringbauer, T.~J. Weinhold, L.~A. Howard, A.~G. White, and M.~R. Vanner.
\newblock ``Generation of mechanical interference fringes by multi-photon
  counting''.
\newblock \href{https://dx.doi.org/10.1088/1367-2630/aabb8d}{New Journal of
  Physics {\bf 20}, 053042}~(2018).

\bibitem{Chegnizadeh2024Quantum}
Mahdi Chegnizadeh, Marco Scigliuzzo, Amir Youssefi, Shingo Kono, Evgenii
  Guzovskii, and Tobias~J. Kippenberg.
\newblock ``Quantum collective motion of macroscopic mechanical oscillators''.
\newblock \href{https://dx.doi.org/10.1126/science.adr8187}{Science {\bf 386},
  1383--1388}~(2024).

\bibitem{Pikovski2012Probing}
Igor Pikovski, Michael~R. Vanner, Markus Aspelmeyer, M.~S. Kim, and
  {\v{C}}aslav Brukner.
\newblock ``Probing {Planck}-scale physics with quantum optics''.
\newblock \href{https://dx.doi.org/10.1038/nphys2262}{Nature Physics {\bf 8},
  393--397}~(2012).

\bibitem{Tobar2024Detecting}
Germain Tobar, Sreenath~K. Manikandan, Thomas Beitel, and Igor Pikovski.
\newblock ``Detecting single gravitons with quantum sensing''.
\newblock \href{https://dx.doi.org/10.1038/s41467-024-51420-8}{Nature
  Communications {\bf 15}, 7229}~(2024).

\bibitem{Kienzler2016Observation}
D.~Kienzler, C.~Fl{\"u}hmann, V.~Negnevitsky, H.-Y. Lo, M.~Marinelli,
  D.~Nadlinger, and J.~P. Home.
\newblock ``Observation of quantum interference between separated mechanical
  oscillator wave packets''.
\newblock \href{https://dx.doi.org/10.1103/PhysRevLett.116.140402}{Physical
  Review Letters {\bf 116}, 140402}~(2016).

\bibitem{Bild2023Schrodinger}
Marius Bild, Matteo Fadel, Yu~Yang, Uwe von L{\"u}pke, Phillip Martin,
  Alessandro Bruno, and Yiwen Chu.
\newblock ``{Schr{\"o}dinger} cat states of a 16-microgram mechanical
  oscillator''.
\newblock \href{https://dx.doi.org/10.1126/science.adf7553}{Science {\bf 380},
  274--278}~(2023).

\bibitem{Frowis2017Lower}
Florian Fr{\"o}wis.
\newblock ``Lower bounds on the size of general {Schr{\"o}dinger}-cat states
  from experimental data''.
\newblock \href{https://dx.doi.org/10.1088/1751-8121/aa5a92}{Journal of Physics
  A: Mathematical and Theoretical {\bf 50}, 114003}~(2017).

\bibitem{Marti2024Quantum}
Stefano Marti, Uwe von L{\"u}pke, Om~Joshi, Yu~Yang, Marius Bild, Andra{\v{z}}
  Omahen, Yiwen Chu, and Matteo Fadel.
\newblock ``Quantum squeezing in a nonlinear mechanical oscillator''.
\newblock \href{https://dx.doi.org/10.1038/s41567-024-02545-6}{Nature Physics
  {\bf 20}, 1448--1453}~(2024).

\bibitem{French1994Interband}
Roger~H. French, David~J. Jones, and Stephen Loughin.
\newblock ``Interband electronic structure of $\alpha$-alumina up to
  2167~{K}''.
\newblock \href{https://dx.doi.org/10.1111/j.1151-2916.1994.tb07009.x}{Journal
  of the American Ceramic Society {\bf 77}, 412--422}~(1994).

\bibitem{Hornberger2012Colloquium}
Klaus Hornberger, Stefan Gerlich, Philipp Haslinger, Stefan Nimmrichter, and
  Markus Arndt.
\newblock ``{Colloquium}: Quantum interference of clusters and molecules''.
\newblock \href{https://dx.doi.org/10.1103/RevModPhys.84.157}{Reviews of Modern
  Physics {\bf 84}, 157--173}~(2012).

\bibitem{Fein2019Quantum}
Yaakov~Y. Fein, Philipp Geyer, Patrick Zwick, Filip Kia{\l}ka, Sebastian
  Pedalino, Marcel Mayor, Stefan Gerlich, and Markus Arndt.
\newblock ``Quantum superposition of molecules beyond 25 {kDa}''.
\newblock \href{https://dx.doi.org/10.1038/s41567-019-0663-9}{Nature Physics
  {\bf 15}, 1242--1245}~(2019).

\bibitem{Pedalino2026Probing}
Sebastian Pedalino, Bruno~E. Ram{\'\i}rez-Galindo, Richard Ferstl, Klaus
  Hornberger, Markus Arndt, and Stefan Gerlich.
\newblock ``Probing quantum mechanics with nanoparticle matter-wave
  interferometry''.
\newblock \href{https://dx.doi.org/10.1038/s41586-025-09917-9}{Nature {\bf
  649}, 866--870}~(2026).

\bibitem{Frowis2017Experimental}
Florian Fr{\"o}wis, Peter~C. Strassmann, Alexey Tiranov, Corentin Gut, Jonathan
  Lavoie, Nicolas Brunner, F{\'e}lix Bussi{\`e}res, Mikael Afzelius, and
  Nicolas Gisin.
\newblock ``Experimental certification of millions of genuinely entangled atoms
  in a solid''.
\newblock \href{https://dx.doi.org/10.1038/s41467-017-00898-6}{Nature
  Communications {\bf 8}, 907}~(2017).

\bibitem{Bose2017Spin}
Sougato Bose, Anupam Mazumdar, Gavin~W. Morley, Hendrik Ulbricht, Marko
  Toro{\v{s}}, Mauro Paternostro, Andrew~A. Geraci, Peter~F. Barker, M.~S. Kim,
  and Gerard Milburn.
\newblock ``Spin entanglement witness for quantum gravity''.
\newblock \href{https://dx.doi.org/10.1103/PhysRevLett.119.240401}{Physical
  Review Letters {\bf 119}, 240401}~(2017).

\bibitem{Clarke2008Superconducting}
John Clarke and Frank~K. Wilhelm.
\newblock ``Superconducting quantum bits''.
\newblock \href{https://dx.doi.org/10.1038/nature07128}{Nature {\bf 453},
  1031--1042}~(2008).

\bibitem{Korsbakken2009Electronic}
Jan~I. Korsbakken, Frank~K. Wilhelm, and K.~Birgitta Whaley.
\newblock ``Electronic structure of superposition states in flux qubits''.
\newblock \href{https://dx.doi.org/10.1088/0031-8949/2009/T137/014022}{Physica
  Scripta {\bf T137}, 014022}~(2009).

\bibitem{vanderWal2000Quantum}
Caspar~H. van~der Wal, A.~C.~J. ter Haar, F.~K. Wilhelm, R.~N. Schouten, C.~J.
  P.~M. Harmans, T.~P. Orlando, Seth Lloyd, and J.~E. Mooij.
\newblock ``Quantum superposition of macroscopic persistent-current states''.
\newblock \href{https://dx.doi.org/10.1126/science.290.5492.773}{Science {\bf
  290}, 773--777}~(2000).

\bibitem{Friedman2000Quantum}
Jonathan~R. Friedman, Vijay Patel, W.~Chen, S.~K. Tolpygo, and J.~E. Lukens.
\newblock ``Quantum superposition of distinct macroscopic states''.
\newblock \href{https://dx.doi.org/10.1038/35017505}{Nature {\bf 406},
  43--46}~(2000).

\bibitem{Korsbakken2010Size}
Jan~I. Korsbakken, Frank~K. Wilhelm, and K.~B. Whaley.
\newblock ``The size of macroscopic superposition states in flux qubits''.
\newblock \href{https://dx.doi.org/10.1209/0295-5075/89/30003}{EPL {\bf 89},
  30003}~(2010).

\bibitem{IbachLuth}
Harald Ibach and Hans L{\"u}th.
\newblock ``Solid-state physics: An introduction to principles of materials
  science''.
\newblock \href{https://dx.doi.org/10.1007/978-3-540-93804-0}{Chapter~10}.
\newblock Springer. Berlin, Heidelberg~(2009).
\newblock 4 edition.

\bibitem{Frowis2018Insufficiency}
Florian Fr{\"o}wis, Benjamin Yadin, and Nicolas Gisin.
\newblock ``Insufficiency of avoided crossings for witnessing large-scale
  quantum coherence in flux qubits''.
\newblock \href{https://dx.doi.org/10.1103/PhysRevA.97.042103}{Physical Review
  A {\bf 97}, 042103}~(2018).

\bibitem{Morris2020Entanglement}
Benjamin Morris, Benjamin Yadin, Matteo Fadel, Tilman Zibold, Philipp
  Treutlein, and Gerardo Adesso.
\newblock ``Entanglement between identical particles is a useful and consistent
  resource''.
\newblock \href{https://dx.doi.org/10.1103/PhysRevX.10.041012}{Physical Review
  X {\bf 10}, 041012}~(2020).

\bibitem{Girolami2014Observable}
Davide Girolami.
\newblock ``Observable measure of quantum coherence in finite dimensional
  systems''.
\newblock \href{https://dx.doi.org/10.1103/PhysRevLett.113.170401}{Physical
  Review Letters {\bf 113}, 170401}~(2014).

\bibitem{Yadin2021Metrological}
Benjamin Yadin, Matteo Fadel, and Manuel Gessner.
\newblock ``Metrological complementarity reveals the {Einstein-Podolsky-Rosen}
  paradox''.
\newblock \href{https://dx.doi.org/10.1038/s41467-021-22353-3}{Nature
  Communications {\bf 12}, 2410}~(2021).

\bibitem{Fan1957Imbedding}
Ky~Fan and Gordon Pall.
\newblock ``Imbedding conditions for {Hermitian} and normal matrices''.
\newblock \href{https://dx.doi.org/10.4153/CJM-1957-036-1}{Canadian Journal of
  Mathematics {\bf 9}, 298--304}~(1957).

\bibitem{Paris2009Quantum}
Matteo G.~A. Paris.
\newblock ``Quantum estimation for quantum technology''.
\newblock \href{https://dx.doi.org/10.1142/S0219749909004839}{International
  Journal of Quantum Information {\bf 7}, 125--137}~(2009).

\bibitem{Graneli1989Vibrational}
Bj{\"o}rn Gran{\'e}li and Ulf Dahlborg.
\newblock ``The vibrational motions in vitreous silica at high temperatures''.
\newblock \href{https://dx.doi.org/10.1016/0022-3093(89)90043-4}{Journal of
  Non-Crystalline Solids {\bf 109}, 295--310}~(1989).

\bibitem{Stickler2015Molecular}
Benjamin~A. Stickler and Klaus Hornberger.
\newblock ``Molecular rotations in matter-wave interferometry''.
\newblock \href{https://dx.doi.org/10.1103/PhysRevA.92.023619}{Physical Review
  A {\bf 92}, 023619}~(2015).

\bibitem{Fein2020Thesis}
Yaakov~Y. Fein.
\newblock ``Long-baseline universal matter-wave interferometry''.
\newblock \href{https://dx.doi.org/10.25365/thesis.62746}{PhD thesis}.
\newblock Universit{\"a}t Wien.
\newblock Vienna~(2020).

\end{thebibliography}

\appendix

\section{Relating extensive size to measure by Nimmrichter and Hornberger} \label{app:nh}
Here, we show a connection between the extensive size and the macroscopicity measure from Ref.~\cite{Nimmrichter2013Macroscopicity}.
Recall that their measure is defined in terms of the ability of an experiment to rule out parameter ranges in a decoherence model minimally modifying quantum mechanics.
This model causes decoherence via random position and momentum shifts in phase space.
In many cases, including with mechanical oscillators (e.g., Ref.~\cite{Schrinski2023Macroscopic}), the relevant part of the model is the random momentum kick, because the superpositions are primarily in position space.
Neglecting position shifts, their master equation then reduces to the form
\begin{align}
    \frac{\dd \rho}{\dd t} & = \frac{1}{\tau_e} \int \dd^3 \vb*{q} \, g(\vb*{q}) \mc{D}[L(\vb*{q})] \rho, \nonumber \\
    g(\vb*{q}) & = \left( \frac{1}{2\pi \sigma_q^2} \right)^{\frac{3}{2}} e^{-\frac{\abs{\vb*{q}}^2}{2 \sigma_q^2}} , \nonumber \\
    L(\vb*{q}) & = \sum_{i=1}^N \frac{m_i}{m_e} e^{i \vb*{q}\cdot\vb*{x}_i / \hbar}.
\end{align}
Here, $\mc{D}[L(\vb*{q})]\rho = L(\vb*{q}) \rho L(\vb*{q})^\dagger - \frac{1}{2}[L(\vb*{q})^\dagger L(\vb*{q}) \rho + \rho L(\vb*{q})^\dagger L(\vb*{q})]$ is the dissipator term corresponding to a shift in momentum by $\vb*{q}$, $m_e$ is the electron mass, $m_i$ and $\vb*{x}_i$ are the mass and position of particle $i$, $\sigma_q$ determines a critical length scale for the decay of position superpositions, and $\tau_e$ is a time-scale parameter setting the overall strength of decoherence.
To stay in the nonrelativistic domain, the critical length scale is restricted to $l_q = \hbar / \sigma_q \gtrsim \SI{10}{\femto \metre}$.
The macroscopicity measure is then the logarithm of the greatest value of $\tau_e$ ruled out by the experiment: $\mu = \log_{10}(\tau_e / \SI{1}{s})$.

Suppose we constrain ourselves to a compact object of extent much less than the critical length scale $l_q$, whose wavefunction is also spread over distances much less than $l_q$.
This recovers the so-called diffusive limit~\cite[page 193]{Nimmrichter2014Macroscopic} -- we can approximate
\begin{align}
    L(\vb*{q}) & \approx \sum_i \frac{m_i}{m_e} \left( \id + i \vb*{q}\cdot \vb*{x}_i / \hbar \right) \nonumber \\
        & = \frac{1}{m_e} \left(M \id + i \vb*{q}\cdot\vb*{Q} / \hbar \right).
\end{align}
The master equation then becomes
\begin{align}
        \frac{\dd \rho}{\dd t} & \approx - \frac{1}{2\tau_e m_e^2 \hbar^2} \int \dd^3 \vb*{q} \, g(\vb*{q}) \sum_{\alpha=x,y,z} q_\alpha^2 [Q_\alpha, [Q_\alpha, \rho]] \nonumber \\
    & = -\frac{\sigma_q^2}{2\tau_e m_e^2 \hbar^2} \sum_{\alpha=x,y,z} [Q_\alpha, [Q_\alpha, \rho]].
\end{align}
Let us further assume that there is one dominant direction, and denote the projection of $\vb*{Q}$ along this direction as $Q$, so
\begin{align} \label{eqn:master_simplified}
    \frac{\dd \rho}{\dd t} & \approx - \frac{\sigma_q^2}{2\tau_e m_e^2 \hbar^2} [Q, [Q,\rho]].
\end{align}

We now claim that a typical decoherence rate $\Gamma$ under this master equation can be related to the extensive size:
\begin{align} \label{eqn:decoherence_rate}
    \Gamma = \frac{2\sigma_q^2 m_u^2 a_0^2}{\tau_e m_e^2 \hbar^2} \extn(\rho, Q).
\end{align}
This is justified via a quantity $\mc{F}_2(\rho,A) := -2 \tr \left( [\rho,A]^2 \right)$ that is similar to the QFI and gives a lower bound $\mc{F}_2(\rho,A) \leq \mc{F}(\rho,A)$~\cite{Girolami2014Observable}.
One possible measure of decoherence speed is the rate of loss of purity, which for the master equation \eqref{eqn:master_simplified} evaluates to~\cite{Lee2011Quantification,Yadin2016General}
\begin{align}
    - \frac{\dd}{\dd t} \tr(\rho^2) = \frac{\sigma_q^2}{2\tau_e m_e^2 \hbar^2} \mc{F}_2(\rho,Q).
\end{align}
The rate $\Gamma$ then comes from the approximation $\mc{F}_2(\rho,Q) \sim \mc{F}(\rho,Q)$.

Suppose the experiment maintains coherence for a time $\tau$; then it has ruled out values of $\Gamma$ larger than $\tau^{-1}$, and thus also $\tau_e$ less than
\begin{align}
    \tau_e \sim \frac{2\tau \sigma_q^2 m_u^2 a_0^2}{m_e^2 \hbar^2} \extn(\rho,Q)
\end{align}
for a given $\sigma_q$.
Then we have
\begin{align} \label{eqn:nh_measure_diffusive}
    \mu & \sim \log_{10} \extn(\rho,Q) + \log_{10}\left(\frac{\tau}{\SI{1}{s}}\right)
    + 2 \log_{10} \left( \frac{\sigma_q a_0}{\hbar} \right) + 2 \log_{10} \left(\frac{m_u}{m_e}\right) + \log_{10} 2 \nonumber \\
        & \sim \log_{10} \extn(\rho,Q) + \log_{10}\left(\frac{\tau}{\SI{1}{s}}\right)
        + 2 \log_{10} \left( \frac{a_0}{l_q} \right) + 6.8.
\end{align}
To obtain a macroscopicity independent of the parameter $l_q$, one maximises over all choices of $l_q$.
In Eq.~\eqref{eqn:nh_measure_diffusive}, increasing $l_q$ always increases $\mu$; however, eventually the diffusive approximation is no longer valid.
For diffraction experiments, the decoherence rate saturates at a maximum when $l_q$ is similar to the interference path separation~\cite{Nimmrichter2013Macroscopicity}.
For the experiments discussed in Sec.~\ref{sec:diffraction}, this is typically $l_q \sim \SI{100}{nm}$.
This coincidentally roughly cancels the final mass term, so
\begin{align}
    \mu \sim \log_{10} \extn(\rho,Q) + \log_{10} \left(\frac{\tau}{\SI{1}{s}}\right).
\end{align}
For smaller $l_q$, outside of the diffusive limit, the decoherence rate is insensitive to the path separation in the superposition, so there is no connection between $\mu$ and $\extn$.

\section{Properties of entangled size} \label{app:entn_properties}
Here, we prove the characteristic properties of the entangled size, in part following the methods in Ref.~\cite{Yadin2017Thesis}.\\

\paragraph{1.~Maximum size:}
By the variance upper bound to the QFI, we have
\begin{align} \label{eqn:max_size_intermediate}
    \entn(\rho, A, \Pi) \leq \frac{\vari (\rho,A)}{\sum_{i=1}^n \vari (\rho, A_{\Pi_i})}.
\end{align}
It is simple to show that
\begin{align}
    \vari (\rho, A) = \sum_{i,j} \mathrm{Cov}(\rho, A_{\Pi_i}, A_{\Pi_j}),
\end{align}
where $\mathrm{Cov}(\rho,A,B) := \frac{1}{2}\tr[\rho (A B + B A)] - \tr[\rho A] \tr[\rho B]$ is the covariance, which satisfies $\mathrm{Cov}(\rho,A,B) \leq \sqrt{\vari (\rho,A) \vari (\rho,B)}$.
Therefore,
\begin{align} \label{eqn:max_size_steps}
    \vari (\rho,A) & \leq \sum_{i,j} \sqrt{\vari (\rho,A_{\Pi_i}) \vari (\rho,A_{\Pi_j})} \nonumber \\
        & = \left( \sum_i \sqrt{\vari (\rho,A_{\Pi_i})} \right)^2 \nonumber \\
        & = n^2 \left( \sum_i \frac{1}{n} \sqrt{\vari (\rho,A_{\Pi_i})} \right)^2 \nonumber \\
        & \leq n^2 \sum_i \frac{1}{n} \vari (\rho,A_{\Pi_i}) \nonumber \\
        & = n \sum_i \vari (\rho,A_{\Pi_i}),
\end{align}
where we have used convexity of the square function in the penultimate line.
The result then follows by inserting this bound into Eq.~\eqref{eqn:max_size_intermediate}.\\

The statement about which states saturate the upper bound has a more technical proof which can be found in Appendix~\ref{app:max_size}.\\

\paragraph{2.~Independent systems:}
The variance and QFI are both additive in the sense that
\begin{align}
    \mc{F}(\rho_1 \ox \rho_2, A_1 + B_2) & = \mc{F}(\rho_1, A_1) + \mc{F}(\rho_2, B_2), \nonumber \\
    \vari (\rho_1 \ox \rho_2, A_1 + B_2) & = \vari (\rho_1, A_1) + \vari (\rho_2, B_2)
\end{align}
when $A_1$ and $B_2$ act on different subsystems $1,2$.
Then the entangled size can be written as
\begin{align}
     \entn(\rho_\alpha \ox \rho_\beta, A_\alpha + A_\beta, \alpha \cup \beta) = \frac{\mc{F}(\rho_\alpha, A_\alpha) + \mc{F}(\rho_\beta, A_\beta)}{4 \sum_{i=1}^{n_\alpha} \vari (\rho_\alpha, A_{\alpha_i}) + 4 \sum_{j=1}^{n_\beta} \vari (\rho_\beta, A_{\beta_j})}  \nonumber \\
     = \frac{\left[ \sum_{i=1}^{n_\alpha} \vari (\rho_\alpha, A_{\alpha_i}) \right] \entn(\rho_\alpha,A_\alpha,\alpha) + \left[ \sum_{j=1}^{n_\beta} \vari (\rho_\beta, A_{\beta_j}) \right] \entn(\rho_\beta,A_\beta,\beta)}{\sum_{i=1}^{n_\alpha} \vari (\rho_\alpha, A_{\alpha_i}) + \sum_{j=1}^{n_\beta} \vari (\rho_\beta, A_{\beta_j})}  \nonumber \\
      \leq \max \left\{ \entn(\rho_\alpha,A_\alpha,\alpha), \, \entn(\rho_\beta,A_\beta,\beta)  \right\}. 
\end{align}\\

\paragraph{3.~Classical mixtures:}
The QFI and variance are respectively convex and concave:
\begin{align}
    \mc{F}(p \rho + [1-p] \sigma, A) & \leq p \mc{F}(\rho, A) + (1-p) \mc{F}(\sigma,A), \nonumber \\
    \vari (p \rho + [1-p] \sigma, A) & \geq p \vari (\rho, A) + (1-p) \vari (\sigma, A),
\end{align}
which implies
\begin{align}
    \entn(p \rho + [1-p] \sigma, A, \Pi) & \leq \frac{p \mc{F}(\rho,A) + (1-p) \mc{F}(\sigma,A)}{4 p \sum_i \vari (\rho,A_{\Pi_i}) + 4(1-p) \sum_i \vari (\sigma,A_{\Pi_i}) } \nonumber \\
        & = \frac{p \left[\sum_i \vari (\rho,A_{\Pi_i})\right] \entn(\rho,A,\Pi) + (1-p)\left[\sum_i \vari (\sigma,A_{\Pi_i})\right]\entn(\sigma,A,\Pi)}{p \sum_i \vari (\rho,A_{\Pi_i}) + (1-p) \sum_i \vari (\sigma,A_{\Pi_i})} \nonumber \\
        & \leq \max \left\{ \entn(\rho,A,\Pi),\,  \entn(\sigma,A,\Pi) \right\}.
\end{align}\\

\paragraph{4.~Multipartite entanglement witness:}
First take a pure state $\ket{\psi} = \ket{\psi_1}\ket{\psi_2}\dots$ such that the $\ket{\psi_j}$ exist on coarse-grained disjoint blocks $\Pi'_j$ each containing no more than $k$ of the subsystems $\Pi_i$.
By properties (1) and (2) above, we have
\begin{align}
    \entn(\psi, A, \Pi) & \leq \max_j \entn(\psi_j, A_{\Pi'_j},\Pi'_j) \nonumber \\
        & \leq k.
\end{align}
A mixed $k$-producible state takes the form $\rho = \sum_r p_r \psi^{(r)} $ of a classical mixture of such pure states.
Property (3) therefore implies $\entn(\rho, A, \Pi) \leq k$.

\section{States with maximum entangled size} \label{app:max_size}
Here, we characterise the states which saturate the maximum possible entangled size $\entn = N$, for a fixed observable and a partition into $N$ subsystems (i.e., $n = N$ in Property~1), with local observables $A_i \equiv A_{\Pi_i}$.

Firstly, Eq.~\eqref{eqn:max_size_intermediate} is saturated if and only if $\pi A \pi \propto \pi$, where $\pi$ is the projector onto the support of $\rho$~\cite[Lemma 1]{Yadin2021Metrological}.
We use the spectral decomposition $\rho = \sum_{k=1}^r p_k \dyad{\psi_k}$, where $r = \mathrm{rank}(\rho)$, so that all $p_k$ are non-zero.
Then $\pi A \pi \propto \pi$ is equivalent to the statement that the restriction of $A$ to the subspace spanned by $\{\ket{\psi_k}\}_{k=1}^r$ is $b \id_r$.
It is easily seen that $b = \ev{A}{\psi_k} \; \forall\, k=1,\dots,r$.
Hence, the $\ket{\psi_k}$ all have the same mean value $\ev{A} = \tr[A \rho]$.
Ref.~\cite{Fan1957Imbedding} gives necessary and sufficient conditions for the existence of a basis in which the $r$-dimensional hermitian matrix $B$ appears as a principal submatrix of the larger $D$-dimensional hermitian matrix $A$.
These are stated in terms of their respective sets of eigenvalues ordered decreasingly: $\alpha_1 \geq \dots \alpha_D$, $\beta_1 \geq \dots \geq \beta_r$.
The necessary and sufficient conditions are
\begin{align} \label{eqn:imbedding_conditions}
    \beta_k \leq \alpha_k \quad \text{and} \quad \beta_{r-k+1} \geq \alpha_{D-k+1} \; \quad \forall \, k = 1,\dots,r.
\end{align}
Here, we take $\beta_k = \ev{A} \; \forall \, k$, and the conditions \eqref{eqn:imbedding_conditions} reduce to
\begin{align} \label{eqn:imbedding2}
    \alpha_{k+D-r} \leq \ev{A} \leq \alpha_k \quad \forall \, k = 1,\dots,r.
\end{align}
In other words, the largest $r$ eigenvalues are at least $\ev{A}$, and the lowest $r$ are at most $\ev{A}$.
Two cases are possible, depending on the rank $r$:
\begin{itemize}
    \item If $r \leq \frac{D+1}{2}$, then $D-r+1 \geq r$, so we must have $\alpha_{D-r+1} \leq \alpha_r$ by the chosen ordering.
    So all we can say from \eqref{eqn:imbedding2} is $\alpha_{D-r+1} \leq \ev{A} \leq \alpha_r$, and any value of $\ev{A}$ is in principle achievable with some $\rho$. 
    \item If $r > \frac{D+1}{2}$, then $D-r+1 < r$ and, since $\alpha_{D-r+1} \leq \ev{A} \leq \alpha_r$, we must have $\ev{A} = \alpha_{D-r+1} = \alpha_{D-r+2} = \dots = \alpha_r$.
    So this eigenvalue has multiplicity of at least $(2r-D)$.
\end{itemize}
Overall, these conditions relating $\ev{A}$ to the eigenvalues of $A$ are \emph{necessary} for $\pi A \pi \propto \pi$.\\

Next, the step $\mathrm{Cov}(\rho,A_i,A_j) \leq \sqrt{\vari (\rho,A_i) \vari (\rho,A_j)}$ used in Eq.~\eqref{eqn:max_size_steps} is due to the Cauchy-Schwarz inequality: letting $\tilde{A}_i = A_i - \ev{A_i}$, and noting that $A_i$ and $A_j$ commute, we have
\begin{align} \label{eqn:cov_inequality}
    \mathrm{Cov}(\rho,A_i,A_j) & = \frac{1}{2}\tr[\rho \{ \tilde{A}_i, \tilde{A}_j \}] \nonumber \\
        & = \tr[\rho \tilde{A}_i \tilde{A}_j] \nonumber \\
        & = \tr[ \sqrt{\rho} \tilde{A}_i \tilde{A}_j \sqrt{\rho} ] \nonumber \\
        & \leq \sqrt{ \tr[\rho \tilde{A}_i^2] \tr[\rho \tilde{A}_j^2] } \nonumber \\
        & = \sqrt{ \vari (\rho,A_i) \vari (\rho,A_j) }.
\end{align}
Equality holds only if $\tilde{A}_i \sqrt{\rho} \propto \tilde{A}_j \sqrt{\rho}$, which is equivalent to $\tilde{A}_i \pi \propto \tilde{A}_j \pi$ -- and this is required to hold for every pair of $i,j$.
Therefore, there are constants $c_i \in \mathbb{C}$ and an operator $C$ such that $\tilde{A}_i \pi = c_i C$.
We first assume that the $c_i$ are not all zero -- if they are, then $\tilde{A}_i \ket{\psi_k} = 0 \, \forall \, i,k$, implying that $\rho$ is a mixture of products of $A_i$ eigenstates, and thus the QFI vanishes.
So let $c_1 \neq 0$ without loss of generality.
We decompose each
\begin{align}
    A_i = \sum_{n,\mu} \alpha^i_n \dyad{\alpha^i_n,\mu}
\end{align}
in terms of its eigenvalues $\alpha^i_n$ with degeneracy labels $\mu$.
Then, from $\tilde{A}_i \ket{\psi_k} = c_i C \ket{\psi_k}$, we obtain
\begin{align} \label{eqn:coeffs}
    (\alpha^i_{n_i} - \ev{A_i}) \braket{\vb*{\alpha_n},\vb*{\mu}}{\psi_k} = c_i \mel{\vb*{\alpha_n},\vb*{\mu}}{C}{\psi_k},
\end{align}
where $\ket{\vb*{\alpha_n},\vb*{\mu}} = \ket{\alpha^1_{n_1},\mu_1}\dots\ket{\alpha^N_{n_N},\mu_N}$.
By considering Eq.~\eqref{eqn:coeffs} with $i=1$ and for arbitrary other $i$, we find
\begin{align}
    \left( c_1[\alpha^i_{n_i} - \ev{A_i}] - c_i [\alpha^1_{n_1} - \ev{A_1}] \right) \braket{\vb*{\alpha_n},\vb*{\mu}}{\psi_k} = (c_1 c_i - c_i c_1) \mel{\vb*{\alpha_n},\vb*{\mu}}{C}{\psi_k} = 0.
\end{align}
This implies that either $\braket{\vb*{\alpha_n},\vb*{\mu}}{\psi_k} = 0$ or
\begin{align}
    \alpha^i_{n_i} - \ev{A_i} = \frac{c_i}{c_1} \left(\alpha^1_{n_1} - \ev{A_1}\right).
\end{align}
Therefore, $n_i$ is determined by $n_1$ (independently of $k$) for all eigenstates of $\rho$ with a non-zero component of $\ket{\vb*{\alpha_n},\vb*{\mu}}$.
Moreover, this index must be such that $\alpha^i_{n_i} = K_i \alpha^1_{n_1} + L_i$ for some constants $K_i,\, L_i$.
Overall, we hence find that the eigenstates of $\rho$ can be written in the GHZ-like form
\begin{align} \label{eqn:ghz_form_intermediate}
    \ket{\psi_k} = \sum_{n, \vb*{\mu}} \psi^k_{n,\vb*{\mu}} \ket{\alpha^1_n,\mu_1} \ket{K_2 \alpha^1_n + L_2,\mu_2} \dots \ket{K_N \alpha^1_n + L_N, \mu_N}.
\end{align}
Note that there is only a single $n$ index. \\

Finally, the last inequality of Eq.~\eqref{eqn:max_size_steps} uses the convexity of the square function.
Explicitly, given a probability distribution $f_i > 0$ and $x_i > 0$,
\begin{align}
    \left(\sum_i f_i \sqrt{x_i} \right)^2 & = \sum_{i,j} f_i f_j \sqrt{x_i x_j} \nonumber \\
        & \leq \sum_{i,j} f_i f_j \left( \frac{x_i+x_j}{2}\right) \nonumber \\
        & = \sum_i f_i x_i,
\end{align}
having used the arithmetic-geometric mean inequality.
This is saturated exactly when all $x_i$ are the same.
So all the single-particle variances $\vari (\rho,A_i)$ must be the same -- implying that the scale factors $K_2,\dots,K_N$ in Eq.~\eqref{eqn:ghz_form_intermediate} are all of unit magnitude.
The fact that $K_i = +1$ is then seen from requiring all covariances to be positive in order to saturate \eqref{eqn:cov_inequality}.
Hence, we have the final form
\begin{align} \label{eqn:ghz_form}
    \ket{\psi_k} = \sum_{n, \vb*{\mu}} \psi^k_{n,\vb*{\mu}} \ket{\alpha^1_n,\mu_1} \ket{\alpha^1_n + L_2,\mu_2} \dots \ket{\alpha^1_n + L_N, \mu_N}.
\end{align}
These states must satisfy the conditions
\begin{subequations}
    \label{eqn:ghz_conditions}
    \begin{align}
        \sum_{n,\vb*{\mu}} (\psi^k_{n,\vb*{\mu}})^* \psi^l_{n,\vb*{\mu}} & = \delta_{k,l} \, , \\
        \sum_{n,\vb*{\mu}} (\psi^k_{n,\vb*{\mu}})^* \psi^l_{n,\vb*{\mu}} \left( N \alpha^1_n + \textstyle\sum_{i=2}^N L_i \right) & = \delta_{k,l} \tr[\rho A]. 
    \end{align}
\end{subequations}
Summarising, we have the following:\\

\paragraph{Maximum size theorem:}
$\entn(\rho,A,\Pi) = N$ if and only if $\rho$ has a set of orthogonal eigenstates of the generalised GHZ form \eqref{eqn:ghz_form}, such that the observable $A$ is constant on the support of $\rho$. 
These conditions are expressed by Eqs.~\eqref{eqn:ghz_conditions}.\\

\paragraph{Qubits:}
Take $N$ qubits and $A_i = Z_i$ (which is without loss of generality, assuming none of them vanish).
Then Eq.~\eqref{eqn:ghz_form} implies that the eigenvectors of $\rho$ are
\begin{align}
    \ket{\psi_k} = \sqrt{1-q_k} \ket{0}^{\ox N} + \sqrt{q_k} e^{i\phi_k} \ket{1}^{\ox N}.
\end{align}
Recall from above that $\ev{A}{\psi_k} = \ev{A} \, \forall k$ -- so the $q_k \equiv q$ are all equal.
Also note from the property $\pi A \pi = \ev{A} \pi$ that $\mel{\psi_k}{A}{\psi_l} = 0$ for $k \neq l$.
Thus, for some pair of eigenvectors $\ket{\psi_0}, \ket{\psi_1}$, we have
\begin{align}
    0 & = \left( \sqrt{1-q} \bra{0}^{\ox N} + \sqrt{q} e^{-i \phi_0} \bra{1}^{\ox N} \right) A \left( \sqrt{1-q} \ket{0}^{\ox N} + \sqrt{q}e^{i\phi_1} \ket{1}^{\ox N} \right) \nonumber \\
        & = N \left( 1 - q - q e^{i(\phi_1-\phi_0)} \right).
\end{align}
This cannot work together with orthogonality $0 = \braket{\psi_0}{\psi_1} = 1-q + q e^{i(\phi_1-\phi_0)}$, which demands $q = \frac{1}{2}$ and $e^{i(\phi_1-\phi_0)} = -1$.
Therefore, the only maximum-size states are the pure GHZ-like states $\sqrt{1-q}\ket{0}^{\ox N} + \sqrt{q}e^{i\phi} \ket{1}^{\ox N}$.
\qed \\

For qutrits and higher, mixed maximum-size states exist.
For example, taking qutrits with $A_i = \dyad{+} + 0 \dyad{0} -\dyad{-}$, the maximum-size states are any classical mixtures of a pair of the form
\begin{align}
    \ket{\psi_0} & = \sqrt{\frac{u}{2}} \left(\ket{+}^{\ox N} + e^{i\phi} \ket{-}^{\ox N} \right) + \sqrt{1-u} e^{i\chi} \ket{0}^{\ox N} \nonumber \\
    \ket{\psi_1} & = \sqrt{\frac{1-u}{2}} \left( \ket{+}^{\ox N} + e^{i\phi} \ket{-}^{\ox N} \right) - \sqrt{u} e^{i\chi} \ket{0}^{\ox N}.
\end{align}
Note that each of these has $\ev{A}{\psi_k} = 0$, while $\vari (\psi_0,A) = u N^2$ and $\vari (\psi_1,A) = (1-u)N^2$.
A mixture $\rho = p \dyad{\psi_0} + (1-p)\dyad{\psi_1}$ has $\vari (\rho,A) = \mc{F}(\rho,A)/4 = [p u + (1-p)(1-u)]N^2$.

\section{Oscillator details} \label{app:oscillator}
We first list some useful facts about the mode quadrature operators.
The mode functions have the orthogonality relations
\begin{align}
    \int \dd^3 \vb*{r} \, w_{\vb*{k}}(\vb*{r}) w_{\vb*{l}}(\vb*{r}) & = \delta_{\vb*{k},\vb*{l}} \, V_{\vb*{k}}, \nonumber  \\
    \sum_{\vb*{k}} \frac{w_{\vb*{k}}(\vb*{r}) w_{\vb*{k}}(\vb*{r'})}{V_{\vb*{k}}} & = \delta^3(\vb*{r}-\vb*{r'}).
\end{align}
The expansions of local position and momentum operators into normal modes are
\begin{align}
    x(\vb*{r}) & = \sum_{\vb*{k}} w_{\vb*{k}}(\vb*{r}) X_{\vb*{k}} , \nonumber \\
    p(\vb*{r}) & = \sum_{\vb*{k}} \frac{w_{\vb*{k}}(\vb*{r})}{V_{\vb*{k}}} P_{\vb*{k}},
\end{align}
and in the other direction,
\begin{align}
    X_{\vb*{k}} & = \int \dd^3 \vb*{r}\, \frac{w_{\vb*{k}}(\vb*{r})}{V_{\vb*{k}}} x(\vb*{r}), \nonumber \\
    P_{\vb*{k}} & = \int \dd^3 \vb*{r}\, w_{\vb*{k}}(\vb*{r}) p(\vb*{r}).
\end{align}
Note that, while $x(\vb*{r})$ is the displacement at position $\vb*{r}$, $p(\vb*{r})$ is the momentum \emph{density}.
Converting from the position observable $x(\vb*{r})$ to its extensive version $q(\vb*{r}) = \varrho x(\vb*{r})$ (for simplicity, assuming constant density), we define $Q_{\vb*{k}} := M_{\vb*{k}} X_{\vb*{k}}$ with the \emph{mode mass} $M_{\vb*{k}} := \varrho V_{\vb*{k}}$.
This leads to
\begin{align}
    q(\vb*{r}) & = \sum_{\vb*{k}} \frac{w_{\vb*{k}}(\vb*{r})}{V_{\vb*{k}}} Q_{\vb*{k}}, \nonumber \\
    Q_{\vb*{k}} & = \int \dd^3 \vb*{r}\, w_{\vb*{k}}(\vb*{r}) q(\vb*{r}).
\end{align}
By normalising the mode functions such that $\max_{\vb*{r}} \abs{w_{\vb*{k}}(\vb*{r})} = 1$, we have for all modes that $V_{\vb*{k}} \leq V$ and $M_{\vb*{k}} \leq M$.
\\

Next, we derive Eq.~\eqref{eqn:modes_variance}.
Using the mode expansion~\eqref{eqn:q_mode_expansion} in the expression Eq.~\eqref{eqn:mode_local_obs}, we have
\begin{align}
    A_i & = \int_{R_i} \dd^3 \vb*{r} \, w_{\vb*{k}}(\vb*{r}) q(\vb*{r}) \nonumber \\
        & = \sum_{\vb*{l}} \int_{R_i} \dd^3 \vb*{r} \, w_{\vb*{k}}(\vb*{r}) w_{\vb*{l}}(\vb*{r}) \frac{Q_{\vb*{l}}}{V_{\vb*{l}}} \nonumber \\
        & = \varrho \sum_{\vb*{l}} \zeta(i,\vb*{k},\vb*{l}) X_{\vb*{l}}.
\end{align}
Since the normal modes are uncorrelated in the state of Eq.~\eqref{eqn:addressed_state}, we find
\begin{align}
        \vari (\rho, A_i) & = \varrho^2 \sum_{\vb*{l}} \zeta(i,\vb*{k},\vb*{l})^2 \vari (\rho, X_{\vb*{l}}) \nonumber \\
        & = \varrho^2 \left[ \zeta(i,\vb*{k},\vb*{k})^2 \vari (\rho_{\vb*{k}},X_{\vb*{k}}) + \sum_{\vb*{l} \neq \vb*{k}} \zeta(i,\vb*{k},\vb*{l})^2 \vari (\gamma_{\vb*{l}}, X_{\vb*{l}}) \right],
\end{align}
and $\vari (\gamma_{\vb*{l}},X_{\vb*{l}}) = \nu_{\vb*{l}} (1 + 2 \bar{n}_{\vb*{l}})$; summing over $i$ results in Eq.~\eqref{eqn:modes_variance}. \\

For the estimates in Sec.~\ref{sec:oscillator_simple}, we first calculate the QFI in a single thermal mode using the general expression~\cite{Paris2009Quantum}
\begin{align}
    \mc{F}(\rho,A) = 2 \sum_{i,j} \frac{(\lambda_i-\lambda_j)^2}{\lambda_i + \lambda_j} \abs{\mel{\psi_i}{A}{\psi_j}}^2,
\end{align}
where $\rho = \sum_i \lambda_i \dyad{\psi_i}$ is the spectral decomposition of $\rho$, and the sum is over all terms such that $\lambda_i + \lambda_j \neq 0$.
A thermal mode can be expanded in the number basis as $\rho = \sum_{n=0}^\infty (1-p)p^n \dyad{n}$, where $p = e^{-\beta \hbar \omega}$, so
\begin{align}
    \mc{F}(\rho,x) = 4(1-p) \sum_{n=0}^\infty \sum_{m=0}^{n-1} \frac{(p^n-p^m)^2}{p^n+p^m} \abs{\mel{m}{x}{n}}^2.
\end{align}
From the matrix element $\mel{m}{x}{n} = \sqrt{\nu n} \delta_{n,m+1}$ with $\nu = \ev{x^2}{0}$, we obtain
\begin{align}
    \mc{F}(\rho,x) & = \frac{4 \nu (1-p)^3}{1+p} \sum_{n=0}^\infty (n+1) p^n \nonumber \\
        & = 4\nu \left( \frac{1-p}{1+p} \right) \nonumber \\
        & = \frac{4\nu}{2\bar{n} + 1} .
\end{align}
Similarly, the variance can be calculated using $\ev{x}{n} = 0$, $\ev{x^2}{n} = \nu(1+ 2n)$, so
\begin{align}
    \vari (\rho,x) = \nu(2 \bar{n} + 1).
\end{align}
An analogous calculation can be done for momentum, giving $\mc{F}(\rho,p) = \hbar^2 / [\nu(2 \bar{n}+1)]$.

For the entangled size with respect to momentum, one has to find the single-particle variance.
Partitioning into atoms, this quantity is not known as directly but can be roughly estimated under the same conditions leading to Eq.~\eqref{eqn:atoms_variance_sum} as $\vari (\rho, U p(\vb*{r_i})) \sim (\hbar / 2 \Delta u)^2$, where $U$ is the effective volume occupied by a single atom.
Expanding the observable $A = P_{\vb*{k}}$ into local parts as $A_i \approx U w_{\vb*{k}}(\vb*{r}_i) p(\vb*{r}_i)$, we then find
\begin{align}
    \sum_i \vari (\rho, A_i) & \sim \int \frac{\dd^3 \vb*{r}}{U} w_{\vb*{k}}(\vb*{r})^2 \left(\frac{\hbar}{2\Delta u} \right)^2 \nonumber \\
        & = \frac{V_{\vb*{k}}}{U} \left(\frac{\hbar}{2 \Delta u}\right)^2 \nonumber \\
        & = N_{\vb*{k}} \left(\frac{\hbar}{2 \Delta u}\right)^2 .
\end{align}
Together with Eq.~\eqref{eqn:oscillator_extn}, this gives
\begin{align}
    \entn(\rho, P_{\vb*{k}}, \Pi) \sim \frac{1}{2 \bar{n} + 1} \cdot \frac{1}{N_{\vb*{k}}} \left(\frac{\Delta u}{\Delta X_{\vb*{k}}^\mathrm{zp}}\right)^2.
\end{align}
This value turns out to be negligible in the examples studied. \\

Here, we give details relating to the experimental estimates listed in Table~\ref{tab:oscillators}.
\begin{itemize}
    \item \textbf{Teufel et al.~(2011)~\cite{Teufel2011Sideband}}:
        A circular drum of ${}^{27} \mathrm{Al}$, radius $R = \SI{7.5}{\micro \meter}$ and thickness $t = \SI{100}{\nano \meter}$, density $\varrho = \SI{2.71e3}{\kilogram \meter^{-3}}$.
        Ref.~\cite{Cartz1955Thermal} gives $\Delta u = \SI{1.7e-11}{\meter}$ at room temperature.
        The fundamental mode is addressed, taking the shape $w_0(r,\theta) = J_0(\alpha_{01} r/R)$ in cylindrical coordinates, with $J_0$ being the zeroth-order Bessel function and $\alpha_{01}$ its smallest zero (note that this is already correctly normalised, as $J_0(0)=1$).
        The mode volume evaluates to $V_0 = \pi R^2 t J_1(\alpha_{01})^2 \approx 0.27 V$.
    \item \textbf{Verhagen et al.~(2012)~\cite{Verhagen2012Quantum}}:
        A silica ($\mathrm{SiO}_2$) torus of minor radius $R_- = \SI{2}{\micro \meter}$ and major radius $R_+ = \SI{15.5}{\micro \meter}$, density $\varrho = \SI{2.65e3}{\kilogram \meter^{-3}}$.
        We average the atomic masses to $\bar{m} = 20m_u$ and take the room-temperature value $\Delta u = \SI{2.5e-11}{\meter}$~\cite{Graneli1989Vibrational}.
        Due to inhomogeneities in the structure, the exact vibrational mode is rather complex, but it is close to a straightforward breathing mode in the outer-radius directions, so we take a constant mode function.
    \item \textbf{Ringbauer et al.~(2018)~\cite{Ringbauer2018Generation}}:
        A square drum of $\mathrm{Si}_3 \mathrm{N}_4$ with side length $L = \SI{1.7}{\milli \meter}$ and thickness $t = \SI{50}{\nano \meter}$, density $\varrho = \SI{3.17e3}{\kilogram \meter^{-3}}$.
        We average the atomic masses to $\bar{m} = 20m_u$; without access to atomic vibrational data, we set $\Delta u \approx \SI{2e-11}{\meter}$.
        The addressed fundamental mode has the shape $w_0(x,y) = \sin(\pi x/L) \times \sin(\pi y/L)$, so $V_0 = V/4$.
    \item \textbf{Chegnizadeh et al.~(2024)~\cite{Chegnizadeh2024Quantum}}:
        $N_\text{osc}=6$ circular drums of ${}^{27} \mathrm{Al}$, each of radius $R = \SI{70}{\micro \meter}$ and thickness $t = \SI{200}{\nano \meter}$.
        The relevant mode $w_{\vb*{k}}(\vb*{r})$ (relative to its centre $\vb*{c_j}$) in each oscillator $j$ is the fundamental, as above for Ref.~\cite{Teufel2011Sideband}, while the collective mode is the sum $w_\text{col}(\vb*{r}) = \sum_{j=1}^{N_\text{osc}} w_{\vb*{k}}(\vb*{r} - \vb*{c}_j)$.
        The collective mode then has mass $N_\text{osc} M_{\vb*{k}}$ and zero-point fluctuations $\Delta X_{\vb*{k}}^\text{zp}/\sqrt{N_\text{osc}}$, where $M_{\vb*{k}}$ and $\Delta X_{\vb*{k}}^\text{zp}$ refer to a single oscillator.
        From Eqs.~\eqref{eqn:oscillator_extn} and \eqref{eqn:oscillator_entn}, relative to a single oscillator, $\extn$ is thus scaled by a factor $N_\text{osc}$, while $\entn$ is unchanged, as would also be the case for $N_\text{osc}$ independent oscillators with the same $\bar{n}$.
    \item \textbf{Rossi et al.~(2024)~\cite{Rossi2024Quantum}}:
        A levitated silica nanoparticle of mass $M = \SI{1.2}{\femto \gram}$, prepared in a squeezed state.
        The relevant motional mode is the CM motion in a harmonic trap (along the $z$-axis of their coordinate system).
        They achieve a ``coherence length'' $\chi = \SI{7.3e-11}{\meter}$, to which we associate $\mc{F}(\rho,X_\text{CM}) = 4 \chi^2$.
        The total position variance is larger, estimated at $\Delta X_\text{CM} = \SI{1.2e-10}{\meter}$ -- which exceeds the molecular vibrations $\Delta u$.
        We then have $\entn(\rho,Q_\text{CM}, \Pi) \approx N \chi^2/(\Delta X_\text{CM}^2 + \Delta u^2) \approx N (\chi / \Delta X_\text{CM})^2$.
\end{itemize}

\section{Diffraction details} \label{app:diffraction}
In the experiment of Ref.~\cite{Fein2019Quantum}, the distribution $p(x)$ is not measured directly.
Instead, just before the detector is another grating whose displacement $s$ in the transverse direction is varied, with the total number passing the grating $n(s)$ being counted.
We show here that, from these counts, one can estimate the probability $R(s)$ of a diffracted particle passing through the final grating.
This probability is related to $p(x)$ by the convolution
\begin{align}
    R(s) = \int \dd x \, p(x) g(x-s),
\end{align}
where $g(x) \in [0,1]$ is the grating transmission function.
We look at the FI of the binary trial $\{R(s), 1-R(s)\}$, with respect to the parameter $s$.
Note that
\begin{align}
    R'(s) & = \frac{\dd}{\dd s} \int \dd x \, p(x+s) g(x) \nonumber \\
        & = \int \dd x \, p'(x+s) g(x) \nonumber \\
        & = \int \dd x \, p'(x) g(x-s),
\end{align}
so a shift in the parameter $s$ is equivalent to a translation of the distribution $p(x)$.
From the FI data processing inequality, it follows that $\mc{F}_\text{cl}[\{R(s),1-R(s)\}] \leq \mc{F}_\text{cl}[p(x)]$.
The binary trial FI is directly evaluated as
\begin{align}
    \mc{F}_\text{cl}[\{R(s),1-R(s)\}] = \frac{R'(s)^2}{R(s)[1-R(s)]}.
\end{align}
In principle, this can be estimated by looking at neighbouring measurement points $R(s)$, $R(s + \delta s)$ and finding the value of $s$ which gives the largest value from a fidelity bound~\cite{Frowis2017Lower}.
However, the probability $R(s)$ is not reported in Ref.~\cite{Fein2019Quantum}, rather a normalised count $n(s) = \mc{N} R(s)$, where the constant $\mc{N}$ is also not given.
We assume that $\mc{N}$ is such that the normalised count averaged over $s$ is unity, $\ev{n} = 1$.
Thus, $1 = \mc{N}\ev{R} = \mc{N} \int \dd x \, p(x) \ev{g} = \mc{N}\ev{g}$, so we have $\mc{N} = 1/\ev{g}$.
Also note that $\ev{g}$ corresponds to the ``open fraction'' of the final grating.

Without directly using the data points in Fig.~2 of Ref.~\cite{Fein2019Quantum}, we instead use a sinusoidal fit to $n(s)$ with mean value $1$, visibility $v$, and wavenumber $k$: $n(s) = 1 + v \sin(ks)$, or
\begin{align}
    R(s) = \ev{g} (1 + v\sin[ks]).
\end{align}
Then we have
\begin{align}
    \mc{F}_\text{cl}[\{R(s),1-R(s)\}] = \frac{\ev{g} v^2 k^2 \cos^2(ks)}{(1 + v\sin[ks]) (1 - \ev{g}[1 + v\sin\{ks\}])}.
\end{align}
A convenient bound (which does not quite maximise the right-hand side) is to take $ks/\pi$ to be an integer, which when inserted into Eq.~\eqref{eqn:diffraction_qfi_bound} gives
\begin{align}
    \mc{F}(\rho_0, Q) \geq \frac{\ev{g}}{1-\ev{g}} (v \hbar k t)^2.
\end{align}

The parameters from Ref.~\cite{Fein2019Quantum} are $\ev{g} = 0.43,\, v=0.25,\, k = 2\pi / \SI{266}{nm}$.
This visibility is attained at an optical grating power of $\SI{1.2}{W}$, where a classical model is found to have negligible visibility.
The distance between the diffraction and final gratings is $\SI{1}{m}$, with an average velocity of about $\SI{260}{m s^{-1}}$, resulting in $\mc{F}(\rho_0, Q) \gtrsim \SI{4.3e-60}{kg^2 m^2}$.
The extensive size is then $\extn(\rho_0,Q) \gtrsim \num{1.4e14}$.
An effective coherence length can be defined as
\begin{align} \label{eqn:coh_length}
    \chi := \frac{\sqrt{\mc{F}(\rho_0,Q)}}{2 M} \geq \sqrt{\frac{\ev{g}}{1-\ev{g}}} \cdot \frac{v \hbar k t}{2M}.
\end{align}
The most abundant molecule in the experiment has $M = \num{26777} m_u$, giving $\chi \gtrsim \SI{20}{nm}$ -- around $10\%$ of the grating period of $\SI{266}{nm}$.

For the entangled size, we need to estimate the single-particle position uncertainty.
The uncertainty in the centre-of-mass frame will include contributions from thermal fluctuations of the shape of the molecule, as well as rotations~\cite{Stickler2015Molecular}.
This cannot be more than the size of the molecule -- a few $\si{nm}$ -- and so is dominated by the centre of mass spread $\Delta X_\text{CM}$.

\begin{figure}[h]
    \centering
    \includegraphics[width=0.6\linewidth]{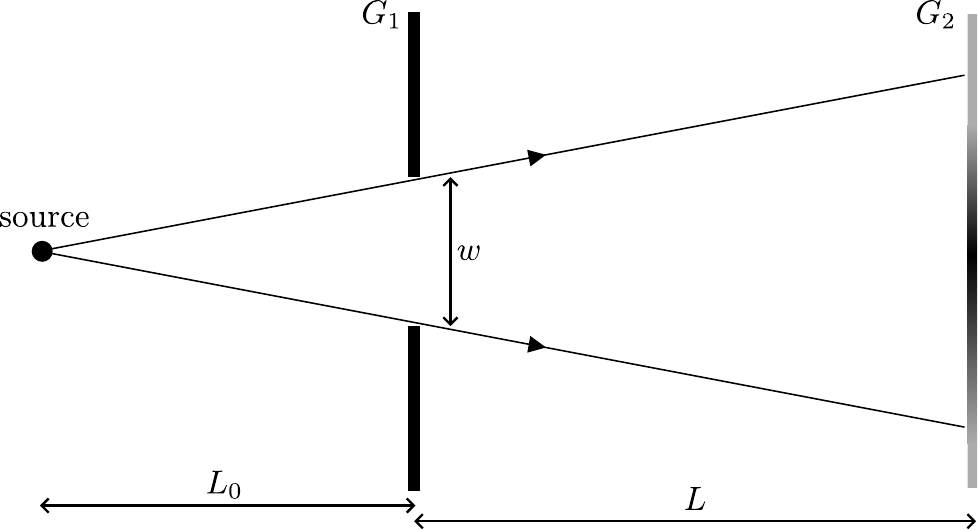}
    \caption{Illustration of the geometry used in estimating the centre-of-mass position variance at the second grating $G_2$. A single slit in the first grating $G_1$ is depicted; given that a molecule passes through this slit, we take a uniform distribution of angles across the width $w$.}
    \label{fig:grating_geometry}
\end{figure}

To estimate $\Delta X_\text{CM}$, consider the set-up of the experiment in Ref.~\cite{Fein2019Quantum}, also detailed in the thesis Ref.~\cite{Fein2020Thesis}.
A low-coherence beam is incident on the first grating $G_1$ whose purpose is to localise the transverse wavefunction such that the particles arrive with significant transverse position coherence at the second grating $G_2$ (see Ref.~\cite{Nimmrichter2014Macroscopic}, Chapter 3).
Due to the low incident coherence at $G_1$, we treat the motion from the source to $G_2$ as ballistic.
The geometry shown in Fig.~\ref{fig:grating_geometry} then implies the following relation between the CM position uncertainties $\Delta X_i$ at the first two gratings:
\begin{align} \label{eqn:xcm_geometry}
    \Delta X_2 = \Delta X_1 \left(1 + \frac{L}{L_0} \right),
\end{align}
where $L_0$ is the distance from the source to $G_1$ and $L$ is the distance from $G_1$ to $G_2$.
Considering a single molecule, its transverse distribution just before $G_1$ should be much larger than its slit width $w$, so passage through a single slit results in an approximately uniform distribution just afterwards.
This has standard deviation $w/\sqrt{12}$ (to be conservative, we take twice this value, $\Delta X_1 = w/\sqrt{3}$); for a grating period $\lambda = \SI{266}{nm}$ and open fraction $\ev{g}=0.43$, we have $w = \ev{g} \lambda = \SI{114}{nm}$.
The distance $L = \SI{1}{\meter}$, and although we do not know $L_0$ precisely, we see from Fig.~37 of Ref.~\cite{Fein2020Thesis} that $L_0 > \SI{0.2}{m}$ (and it is unlikely to be more than around $\SI{1}{m}$).
Inserting parameters into Eq.~\eqref{eqn:xcm_geometry}, we obtain $\Delta X_\text{CM} = \Delta X_2 < \SI{396}{nm}$.

Writing the entangled size (with respect to a partition into atoms) as
\begin{align}
    \entn(\rho_0,Q,\Pi) = N \cdot \left(\frac{\chi}{\Delta X_\text{CM}} \right)^2,
\end{align}
we finally have $\entn \gtrsim 2000 \cdot (20/396)^2 \approx 5$.
If we instead take $L_0 = \SI{1}{m}$, this value can be improved to around 45.

The same procedure can be applied to the experiment of Ref.~\cite{Pedalino2026Probing}, in which all three gratings are optical photodepletion gratings of period $d = \SI{133}{nm}$ separated by $L = \SI{0.983}{m}$, with mean velocity $\SI{160}{m s^{-1}}$ (so $t \approx \SI{6.1}{ms}$) and maximal visibility $v = 0.10$ for clusters of mean mass $\num{1.72e5} m_u$.
A photodepletion grating has transmission function $g(x) = \exp[-n_0 \cos^2(\pi x/d)]$, where $n_0$ is the mean number of absorbed photons in an antinode, so that $\ev{g} = e^{-n_0/2} I_0(n_0/2)$.
Using the reported grating powers, beam waist and ionization cross-section~\cite{Pedalino2026Probing}, we estimate $n_0 \approx 75$ and $83$ for $G_1$ and $G_3$, respectively, hence $\ev{g} \approx 0.06$ for $G_3$.
This results in $\mc{F}(\rho_0,Q) \gtrsim \SI{6.2e-61}{kg^2 m^2}$, i.e., $\extn(\rho_0,Q) \gtrsim \num{2e13}$ and $\chi \gtrsim \SI{1.4}{nm}$.
Near its nodes, $G_1$ transmits an approximately Gaussian window of standard deviation $d/(\pi\sqrt{2n_0}) \approx \SI{3.5}{nm}$; conservatively taking twice this value and $L_0 = \SI{1}{m}$ in Eq.~\eqref{eqn:xcm_geometry}, we obtain $\Delta X_\text{CM} \approx \SI{14}{nm}$ and thus $\entn \approx \num{7.5e3} \cdot (1.4/14)^2 \sim 80$.

\vspace{5mm}

\end{document}